\documentclass[showpacs,preprintnumbers,twocolumn,aps,prl]{revtex4}

\usepackage{amsmath,amssymb,amsthm}
\usepackage{graphicx}
\usepackage{tabularx}
\usepackage{paralist}
\usepackage{epstopdf}
\usepackage{bbold}
\usepackage{braket}
\usepackage{pgfgantt}
\usepackage{pdfpages}
\usepackage{lmodern}

\usepackage{color}
\usepackage{bm}

\usepackage{lpic}
\usepackage[symbol]{footmisc}
\usepackage{tikz, tikzscale}
\usetikzlibrary{matrix, fit}
\usetikzlibrary{decorations.pathreplacing,angles,quotes}
\usetikzlibrary{arrows,fit, quotes,arrows.meta, spy, calc, decorations.markings, shapes.misc, positioning, patterns, shadings, decorations.shapes}
\makeatletter
\newcommand\footnoteref[1]{\protected@xdef\@thefnmark{\ref{#1}}\@footnotemark}
\makeatother

\newcommand{\mbpold}[1]{{\color{black} #1}}

\newcommand{\bene}[1]{{\color{black} #1}}

\tikzset{
	ncbar angle/.initial=90,
	ncbar/.style={
		to path=(\tikztostart)
		-- ($(\tikztostart)!#1!\pgfkeysvalueof{/tikz/ncbar angle}:(\tikztotarget)$)
		-- ($(\tikztotarget)!($(\tikztostart)!#1!\pgfkeysvalueof{/tikz/ncbar angle}:(\tikztotarget)$)!\pgfkeysvalueof{/tikz/ncbar angle}:(\tikztostart)$)
		-- (\tikztotarget)
	},
	ncbar/.default=0.5cm,
}

\tikzset{square left brace/.style={ncbar=0.5cm}}
\tikzset{square right brace/.style={ncbar=-0.5cm}}

\tikzset{round left paren/.style={ncbar=0.5cm,out=120,in=-120}}
\tikzset{round right paren/.style={ncbar=0.5cm,out=60,in=-60}}

\newlength{\hatchspread}%
\newlength{\hatchthickness}%
\newlength{\hatchshift}%
\newcommand{\hatchcolor}{}%
\tikzset{hatchspread/.code={\setlength{\hatchspread}{#1}},
	hatchthickness/.code={\setlength{\hatchthickness}{#1}},
	hatchshift/.code={\setlength{\hatchshift}{#1}},
	hatchcolor/.code={\renewcommand{\hatchcolor}{#1}}}
\tikzset{hatchspread=3pt,
	hatchthickness=0.4pt,
	hatchshift=0pt,
	hatchcolor=black!08}
\pgfdeclarepatternformonly[\hatchspread,\hatchthickness,\hatchshift,\hatchcolor]
{custom north west lines}
{\pgfqpoint{\dimexpr-2\hatchthickness}{\dimexpr-2\hatchthickness}}
{\pgfqpoint{\dimexpr\hatchspread+2\hatchthickness}{\dimexpr\hatchspread+2\hatchthickness}}
{\pgfqpoint{\dimexpr\hatchspread}{\dimexpr\hatchspread}}
{
	\pgfsetlinewidth{\hatchthickness}
	\pgfpathmoveto{\pgfqpoint{0pt}{\dimexpr\hatchspread+\hatchshift}}
	\pgfpathlineto{\pgfqpoint{\dimexpr\hatchspread+0.15pt+\hatchshift}{-0.15pt}}
	\ifdim \hatchshift > 0pt
	\pgfpathmoveto{\pgfqpoint{0pt}{\hatchshift}}
	\pgfpathlineto{\pgfqpoint{\dimexpr0.15pt+\hatchshift}{-0.15pt}}
	\fi
	\pgfsetstrokecolor{\hatchcolor}
	\pgfusepath{stroke}
}

\tikzset{
	ncbar angle/.initial=90,
	ncbar/.style={
		to path=(\tikztostart)
		-- ($(\tikztostart)!#1!\pgfkeysvalueof{/tikz/ncbar angle}:(\tikztotarget)$)
		-- ($(\tikztotarget)!($(\tikztostart)!#1!\pgfkeysvalueof{/tikz/ncbar angle}:(\tikztotarget)$)!\pgfkeysvalueof{/tikz/ncbar angle}:(\tikztostart)$)
		-- (\tikztotarget)
	},
	ncbar/.default=0.5cm,
}

\tikzset{square left brace/.style={ncbar=0.5cm}}
\tikzset{square right brace/.style={ncbar=-0.5cm}}

\tikzset{round left paren/.style={ncbar=0.2cm,out=110,in=-110}}
\tikzset{round right paren/.style={ncbar=0.5cm,out=70,in=-70}}

\tikzset{decorate sep/.style 2 args=
	{decorate,decoration={shape backgrounds,shape=circle,shape size=#1,shape sep=#2}}}

\begin{document}
\title{Robust optical polarisation of nuclear spin baths using Hamiltonian engineering of NV centre quantum dynamics}
\author{Ilai Schwartz$^{1,3,\dagger}$\footnote{\label{note1}qiong.chen@uni-ulm.de, ilai.schwartz@uni-ulm.de, martin.plenio@uni-ulm.de}, Jochen Scheuer$^{2,\dagger}$, Benedikt Tratzmiller$^{1,\dagger}$, Samuel M{\"u}ller$^{2}$, Qiong Chen$^{1}$\footnoteref{note1}, Ish Dhand$^{1}$, Zhenyu Wang$^{1}$,  Christoph M{\"u}ller$^{3}$, Boris Naydenov$^{2}$, Fedor Jelezko$^{2}$ and Martin B. Plenio$^{1}$\footnoteref{note1}}
\affiliation{$^{1}$ Institut f\"{u}r Theoretische Physik und IQST, Albert-Einstein-Allee 11, Universit\"{a}t Ulm, 89081 Ulm, Germany, \\
$^{2}$ Institut f\"{u}r Quantenoptik und IQST, Universit{\"a}t Ulm, 89081 Ulm, Germany,\\
$^{3}$ NVision Imaging Technologies GmbH, 89134 Blaustein, Germany \\
$^{\dagger}$ These authors contributed equally to this work}

\begin{abstract}

Dynamical nuclear polarisation (DNP) is an important technique that uses polarisation transfer
from electron to nuclear spins to achieve nuclear hyperpolarisation. As the electron spin of the nitrogen vacancy (NV) centres in diamond can be optically initialised nearly perfectly even at room temperature and ambient conditions, new opportunities become possible by the combination of efficient DNP with optically polarised NV centres. Among such applications are nanoscale nuclear magnetic resonance spectroscopy of liquids, hyperpolarised nanodiamonds as MRI contrast agents as well as the initialisation of nuclear spin based diamond
quantum simulators.
Current realisations of DNP perform the polarisation transfer by achieving energetic resonance between electrons and nuclei via carefully tuned
microwave fields or by using quasi-adiabatic sweep-based schemes across resonance points. The
former limits robustness against control errors while the latter limits polarisation rates, making the realisation of the applications extremely challenging.
Here we introduce the concept of Hamiltonian engineering by pulse sequences and use it for
the systematic design of polarisation sequences that are simultaneously robust and fast.
We derive sequences theoretically and demonstrate experimentally that they are capable of
efficient polarisation transfer from an optically polarised nitrogen-vacancy centre in
diamond to the surrounding $^{13}$C nuclear spin bath even in the presence of control
errors, making it an ideal tool for the realisation of the above NV centre based applications.

\end{abstract}
\maketitle

\textit{Introduction ---} A key challenge in the quantum manipulation and detection of small nuclear spin ensembles is their minute level of polarisation at thermal equilibrium which is of the order of $10^{-5}$ for a magnetic field of 2 Tesla at room temperature. Overcoming this challenge holds the key for the realisation of quantum applications ranging from quantum simulators to nanoscale NMR devices. These will be turned from lab demonstrations to realistic applications by polarisation schemes that enable their efficient initialisation and read-out.
An important breakthrough in this respect has been the realisation that the electron spin of the nitrogen vacancy (NV) centres in diamond (as well as colour centres in silicon carbide and photo-excited triplet state molecules) can be optically initialised nearly perfectly even at room temperature and ambient conditions. allowing for rapid electron spin polarisation to be generated optically or chemically far above thermal equilibrium and subsequently transferred to surrounding  nuclei~\cite{london2013,scheuer2016optically,alvarez2015local,king2015room,pagliero2014recursive,drake2015influence,acebal2018toward}.
These systems present a unique opportunity for polarising and initialising spin baths and unlocking the potential of nanoscale applications including the initialisation of quantum simulators based on nuclear spin arrays in diamonds~\cite{cai2013}, magnetic resonance imaging (MRI) tracers via diamond nanoparticles~\cite{rej2015hyperpolarized,chen2015optical, kwiatkowski2018direct, bretschneider2016potential, waddington2017phase} and the enhancement of nanoscale nuclear magnetic resonance (NMR) using NV spin ensembles~\cite{schwartz2017blueprint, bucher2017high, schmitt2017submillihertz, acebal2018toward}. 


However, when trying to achieve nuclear hyperpolarisation via NV centres or similar systems (e.g. SiC colour centres, photo-excited triplet state molecules), we are faced with
two key challenges. First, optically polarisable electron spins are typically higher spin systems,
with $S\geq 1$. The lattice-oriented zero-field splitting in the presence of external magnetic
fields \mbpold{and disorder} results in a large spectral range of the electron spin resonance,
hindering effective polarisation transfer to the surrounding nuclear spins. Second, many of
these systems exhibit either fast electron spin relaxation (e.g. photo-excited triplet
states~\cite{tateishi2014room}), requiring fast polarisation transfer, or weak hyperfine
coupling to the nuclear spins (e.g. NV centres in diamonds, especially to external nuclear spins) limiting polarisation transfer
\mbpold{rates}. Overcoming this combination of challenges presents a daunting task, as it
requires a polarisation transfer scheme which (1) works for a large spectral range of the
electronic system and (2) produces a fast and efficient polarisation transfer from electron
to nuclear spins.

Many DNP protocols have been developed and applied over the past several decades, starting
from continuous MW irradiation~\cite{abragam1958new,wenckebach2008solid} to more efficient
pulsed schemes~\cite{henstra1988nuclear,henstra1988enhanced,weis2000solid}. A common theme
in all protocols that are effective for low electron spin concentration is the use of a
long microwave (MW) pulse to match the Larmor frequency of the nuclear spins to the electronic
Rabi rotation in the frame of reference of the MW drive, which is well-known as a Hartmann-Hahn
(H-H) resonance~\cite{hartmann1962nuclear}. \mbpold{Unfortunately, owing to the weak electron-nuclear
interaction, even a small detuning from this resonance inhibits the polarisation transfer
which renders such schemes strongly dependent on the intensity and frequency of the MW drive
as well as the orientation and transition frequency of the NV centre}. While modified protocols
such as the integrated solid effect (ISE)~\cite{henstra1988enhanced,chen2015optical} improve
the robustness across larger spectral ranges, this comes at the expense of a significantly
slower polarisation transfer.
Thus, devising a DNP protocol which is both robust and fast \mbpold{has} remained an unmet
challenge.

Here we present a new approach for performing dynamic nuclear polarisation, termed PulsePol,
which combines fast polarisation transfer with remarkable robustness against a broad range
of experimental imperfections including power and detuning fluctuations. In sharp contrast
to the schemes described above, which allow for polarisation transfer only during the pulses,
this is achieved by the design of sequences of short pulses (much shorter than the nuclear
Larmor period), with extended waiting periods between the pulses, that refocus the electron-nuclear
interaction such that polarisation transfer is achieved through the accumulated dynamics
between pulses.
As we will demonstrate both theoretically and experimentally, this approach provides inherent
robustness to MW errors and electron spectral width and a high degree of flexibility which
allows to take full advantage of pulse optimisation methods developed in NMR \cite{shaka1987symmetric,wimperis1994BB1,skinner2012new,spindler2012shaped,kobzar2012exploring}
to tailor the polarisation to the specific challenges of the experimental setup.
	
We start by describing the theoretical framework of Hamiltonian engineering and derive
the PulsePol sequence whose robustness against detuning and pulse length/strength errors
we establish theoretically. We then proceed to confirm experimentally the efficiency
of the protocol for a single NV centre in diamond surrounded by ${}^{13}$C nuclei, and
compare the performance of the PulsePol protocol to state-of-the-art DNP schemes. Finally,
we discuss several applications of our protocol.

\textit{Theoretical framework ---} For simplicity we consider a system of a single electron
spin $\vec{S}$ ($S=\frac{1}{2}$ for simplicity, later realised by the $\ket{m=0}$ and
$\ket{m=-1}$ levels of the NV centre) coupled to a single nuclear spin $\vec{I}$, subject
to intermediately applied microwave pulses which are used for control of the electron spin
\begin{equation} \label{H}
    H = \omega_S S_z + \omega_I I_z + \vec{S} \mathcal{A} \vec{I} + H_d,
\end{equation}
where $\omega_S$($\omega_I$) denotes the electron (nuclear) Larmor frequency, $\mathcal{A} $
the hyperfine tensor. $H_d = 2 \Omega(t) S_x \cos(\omega_{MW} t + \varphi)$, in which
$\omega_{MW}$ is the microwave frequency, $\varphi$ its phase, and the Rabi frequency
$\Omega(t)$ takes the value $\Omega_0$ when the microwave is on, and $0$ otherwise.

\mbpold{For the realisation of a robust and efficient interaction between the NV centre and
the surrounding nuclear spins, one needs to achieve simultaneously the decoupling of the
NV-electron spin from environmental noise while refocussing the desired interaction with
the nuclear bath. This can be achieved by dynamical decoupling (DD) protocols
\cite{yang2011preserving,souza2012robust} such as Carr-Purcell-Meiboom-Gill(CPMG)
\cite{Carr:1954:630,MeiboomRSI1958} or the XY pulse family~\cite{MaudsleyJMR1986} where
equally spaced pulses separated by a time $\tau$ such that only interactions \mbpold{with
nuclear spins precessing at a} frequency $\omega_I = n\pi/\tau$ is preserved.}

\begin{figure}[t!]
	
\includegraphics[width=\linewidth]{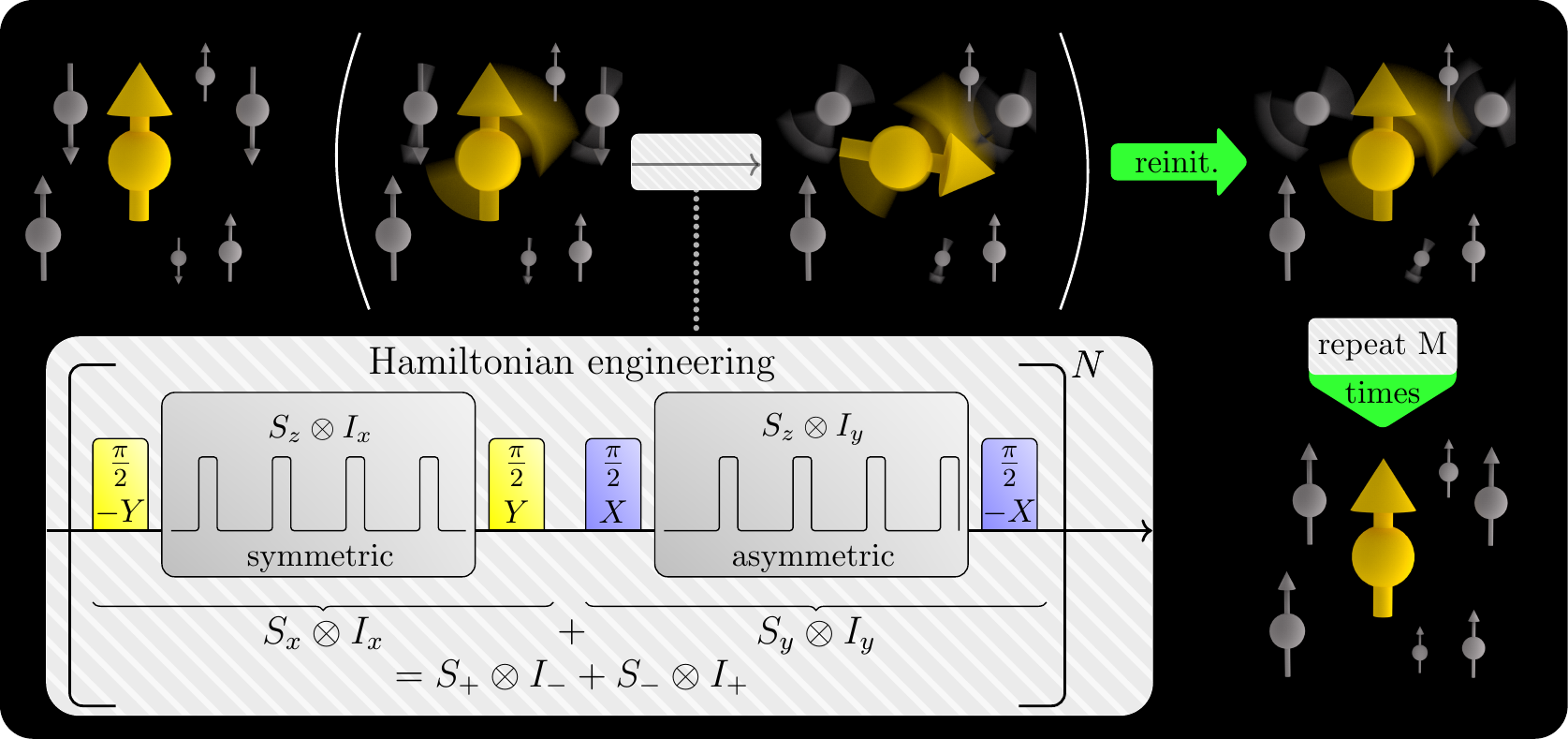}

\caption{
	Polarisation of a bath of nuclear spins (grey) by an electron spin (yellow): The initially unpolarised bath and the polarised electron spin (left top) evolve according to a engineered Hamiltonian that describes a flip-flop interaction, the electron spin is periodically reinitialised. After several repetitions, the nuclear spin bath is polarised (right bottom).
	Illustration of effective Hamiltonian engineering (grey box): A combination of standard symmetric and asymmetric dynamical decoupling sequences with the effective Hamiltonians $H \propto S_z \otimes I_{x/y}$ can be modified to a flip-flop Hamiltonian $H \propto S_x \otimes I_x + S_y \otimes I_y$ by introducing basis changes with $\pi/2$ pulses.}	
\label{sequence}
\end{figure}

When these DD sequences achieve resonance with a nuclear spin, the resulting effective Hamiltonian
that describes the time evolution is, in a suitable electronic basis, of the form $H_\mathrm{eff}
= \alpha A_xS_xI_x$ where $A_x$ denotes the $x$ component of the hyperfine vector, and $\alpha<1$
is a constant determined by the filter function generated by the pulse sequence \cite{kolkowitz2012sensing,taminiau2012detection,casanova2015robust}.
$H_\mathrm{eff}$ does not transfer polarisation and the concept of pulsed polarisation is
to engineer this Hamiltonian to produce the desired flip-flop dynamics. To this end, for a
short time interval $\Delta t\ll \omega_I^{-1}$ we apply a time-evolution according to $H_\mathrm{eff} =
\alpha A_xS_xI_x$ and in a subsequent interval of length $\Delta t$ we let the system follow
a time evolution according to $H_\mathrm{eff} = \alpha A_xS_yI_y$. This is achieved by mapping
both the electronic basis by a MW pulse and the nuclear spin basis by a suitable time delay
from x to y (see Fig.~\ref{sequence}). Repeating this sequence, yields a time evolution that
is governed, for times exceeding $\omega_I^{-1}$ and whenever the coupling strength satisfies
$A_x\ll \omega_I$, by the effective flip-flop Hamiltonian
\begin{equation} \label{flipflop}
    H_\text{avg} = -\frac{\alpha A_x}{4} \left( S_+  I_- + S_-  I_+  \right)
\end{equation}
which is the sum of $H_\mathrm{eff} = \alpha A_xS_xI_x$ and $H_\mathrm{eff} =
\alpha A_xS_yI_y$.
We would like to stress that the rapid switching in intervals $\Delta t$ represents a crucial deviation
from schemes that aim to achieve engineered electron-nuclear SWAP gates after a time $\pi A_x^{-1}$ by applying
first an evolution according to $\exp(-iA_xS_{{x}}I_x t/2)$ for half the total interaction time
followed by $\exp(-iA_xS_{{y}}I_y t/2)$ for the second half \cite{taminiau2014universal,wang2017delayed}
as this achieves a SWAP gate only at $t=\pi A_x^{-1}$ while in between the evolution may include both
flip-flip and flip-flop terms. Achieving flip-flop dynamics on the time-scale of the Larmor frequency $\omega_I$ is
crucial for DNP applications, especially those in which the nuclei are in rapid motion as standard SWAP
pulse sequences are slow and are not suitable for ensembles of electron spins coupled to nuclear spin
baths due to coupling strength variations. Furthermore, any loss of coherence faster than the interaction
time, due to noisy environment or dynamics on fast time-scales (e.g. molecular motion), would completely
destroy any polarisation transfer. These effects are mitigated when the average dynamics are achieved on
the nuclear Larmor time-scale, which are very fast.

However, one challenge remains, as the basic scheme using the rapid alternation between $A_xS_{{x}}I_x$
and $A_xS_{{y}}I_y$ on the time scale of the Larmor frequency, and other simple variants, come at the
expense of enhanced sensitivity to pulse errors and detunings because the change between $S_x$ and $S_y$
dynamics typically introduce unbalanced $\pi/2$ pulses, not part of standard dynamical decoupling
sequences.

Thus, no robust pulse sequence or tailoring of DD sequence is known to produce the fast flip-flop dynamics
on very short time scales, regardless of the coupling strength while satisfying (i) that detuning errors
accumulated during the free evolution are cancelled (e.g. by $\pi$ refocussing~\cite{suppinfo}), and preferably
also decoupled from unwanted noise and fluctuations, (ii) that variable pulse lengths and detuning and Rabi
frequency errors, are canceled at least to the first order.

We have constructed such a sequence, termed PulsePol and depicted in figure~\ref{PulsePol_sim}(a),
analytically from components chosen to satisfy the above requirements (see~\cite{suppinfo} for a
detailed discussion). PulsePol achieves polarisation transfer for a choice of pulse
spacing
\begin{equation}
    \tau = \frac{n\pi}{\omega_I},
\end{equation}
for odd $n$ \cite{fn1} with the strongest coupling occurring for $n=3$, where $\alpha = \frac{2}{3\pi}
\left(2+\sqrt{2}\right)$ and yields an effective flip-flop Hamiltonian for times longer than
$\omega_I^{-1}$. Full polarisation transfer between NV and nuclei is found after a time $t = 2N\tau =
4\pi/\alpha A_x$ where $N$ is the number of basic pulse cycles (see figure \ref{PulsePol_sim}a).
As a consequence PulsePol achieves polarisation transfer in a time scale that is only 28\% slower
than that achieved with Hartmann-Hahn coherent transfer under optimal conditions (see
figure~\ref{PulsePol_sim}(b)), and significantly faster than other polarisation transfer schemes
(e.g. ISE, solid effect).
Furthermore, PulsePol is robust to errors as it has several inherent advantages that enhance
its robustness:
	
	\begin{figure}
		\center

\includegraphics[width=\linewidth]{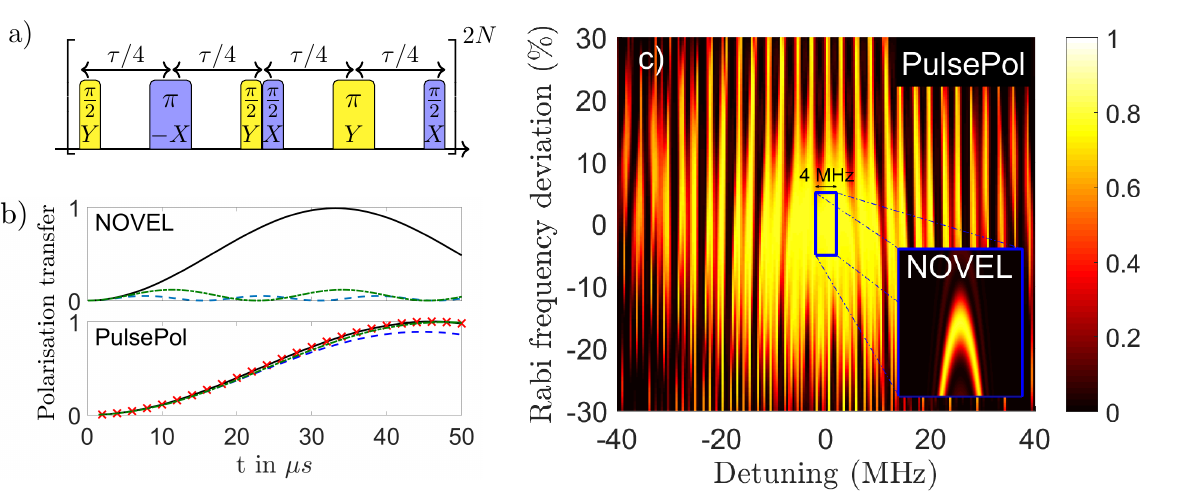}

		
\caption{PulsePol sequence (a) and robustness compared to NOVEL. (b) The
 polarisation transfer to a single nuclear spin for NOVEL
(upper graph) under perfect conditions (black line). A $\Delta= \bene{(2\pi)} 0.5$MHz detuning
error (dashed blue line) and $2\%$ Rabi frequency error (dash-dot green line) drastically reduce polarisation transfer.
For PulsePol (lower graph) under perfect conditions (black line) the transfer is
28\% slower as predicted by the effective Hamiltonian of Eq.~\ref{flipflop}
(red x). Detuning errors  ${\Delta}= 0.1 {\Omega_0} = \bene{(2\pi)}5$MHz (dashed blue line)
and Rabi frequency errors ${\delta\Omega}=0.1 {\Omega_0} = \bene{(2\pi)}5$MHz (dash-dot green
line) have a very small effect.
(c) Considering a system of one electron spin and 5 nuclear spins, for the
same parameters $\omega_I=\bene{(2\pi)}2\;\text{MHz}$ and $\Omega_0
= \bene{(2\pi)}50 \;\text{MHz}$, polarisation transfer vs. $\Delta$ and $\delta\Omega/\Omega_0$
for the PulsePol sequence, with a comparison to a NOVEL sequence for its
relevant detuning values $|\Delta|<\bene{(2\pi)}2 $MHz in the insert. The graphs result from averaging over 100 realisations of the locations of the 5 closest nuclear spins to the NV centre electron spin on a carbon lattice, and for PulsePol a resonance shift of 2.5\% and corresponding phase errors were used, see supplementary information \cite{suppinfo}.}
		\label{PulsePol_sim}
	\end{figure}

\begin{itemize}
    \item Strong pulses with amplitude $\Omega_0$, without the requirement of matching the
	Hartmann-Hahn condition, are robust to detuning satisfying $\Delta \ll \Omega_0$
	\item The PulsePol sequence cancels second order errors in the pulse strength and first
        order in the detuning~\cite{suppinfo}, making it robust to noise and small pulse
        imperfections, similar to dynamical decoupling sequences
	\item The individual pulses can be optimised by numerous methods developed in NMR, including
		composite pulses~\cite{wimperis1994BB1,shaka1987symmetric,alway2007arbitrary} and shaped
        pulses~\cite{kobzar2012exploring,spindler2012shaped}
	\item 
        Phase errors in the applied pulses can be corrected to first order
        by a corresponding shift in the resonance condition with no loss
        of robustness to detuning and Rabi frequency errors (see SI for
        further details).
	\end{itemize}

For demonstrating the robustness of the PulsePol sequence, we introduce errors into the pulses and free evolutions,
with the PulsePol evolution during the pulses described as
\begin{equation}
    \begin{split}
    U_{\theta, \pm \text{X/}\pm \text{Y}} = \exp \biggl[-i \frac{\theta}{\Omega_0} \Bigl( \Delta S_z \pm \tilde \Omega_0 S_{X/Y} + \\
    \sum_j\omega_I I^j_z + S_z \left(A^j_x I^j_x + A^j_z I^j_z  \right)  \Bigr)\biggr],
    \end{split}
\end{equation}
including the detuning (frequency mismatch) $\Delta$ that is also present during
every free evolution and accounting for deviations in pulse strength/length of
the pulses by a Rabi frequency error $\delta\Omega = \Omega_0 - \tilde\Omega_0$.
As the pulse duration is finite, it is subtracted from the waiting time between
the pulses. Since the PulsePol sequence corrects the errors after \mbpold{two cycles
lasting a total of \mbpold{$2\tau$}}, fluctuations in $\Omega_0$ or $\omega_S$ (and
hence $\Delta$) which occur on a slower timescale (e.g. for X-band frequencies and
$^{13}$C or $^{1}$H \mbpold{this implies $2\tau < 1\, \mathrm{\mu s}$)} will have a
similar effect to a constant error, thus the following described robustness holds
also for magnetic field or microwave inhomogeneities.

Fig.~\ref{PulsePol_sim}(b) demonstrates the effect of pulse errors by showing the
evolution of the polarisation of a single nuclear spins as quantified by $2\langle
I_z\rangle$ near an initially fully polarised electron spin, when applying the
PulsePol sequence and the NOVEL sequence \cite{henstra1988nuclear}. The chosen
parameters are typical for an NV centre spin in a diamond surrounded by $^{13}$C
nuclear spins. Compared to NOVEL, where small errors can almost completely eliminate
the polarisation transfer, PulsePol is only slightly affected by these errors.

For a detailed characterisation of the robustness, we consider a system of nuclear
spins coupled to an electron spin in a diamond. Due to the 1.1\% natural abundance
of $^{13}$C isotopes, the interaction is dominated by the closest nuclear spins and
we restrict attention of the dynamics of the 5 closest nuclear spins to the NV centre
electron spin on the lattice. The resulting polarisation transfer efficiency from
the fully polarised NV centre is quantified by $2\sum_i [\braket{I^i_z(t)} - \braket{I^i_z(0)}]$
for different values of $\Delta$, $\delta\Omega$ and averaged over 100 nuclear spins
configurations is shown in Fig.~\ref{PulsePol_sim}(c). As can be seen, efficient
polarisation transfer can be achieved in a $\bene{(2\pi)}60$MHz spectral range for
the \mbpold{microwave driving strength of} $\Omega_0 = \bene{(2\pi)}50$MHz. Note that
the periodic vertical polarisation resonance lines in Fig.~\ref{PulsePol_sim}(c) are
due to the detuning matching a resonance condition \bene{$\Delta \tau/4 = k\pi$ for
an integer $k$} during the free evolution.

\begin{figure*}
	\center

\includegraphics[width=\linewidth]{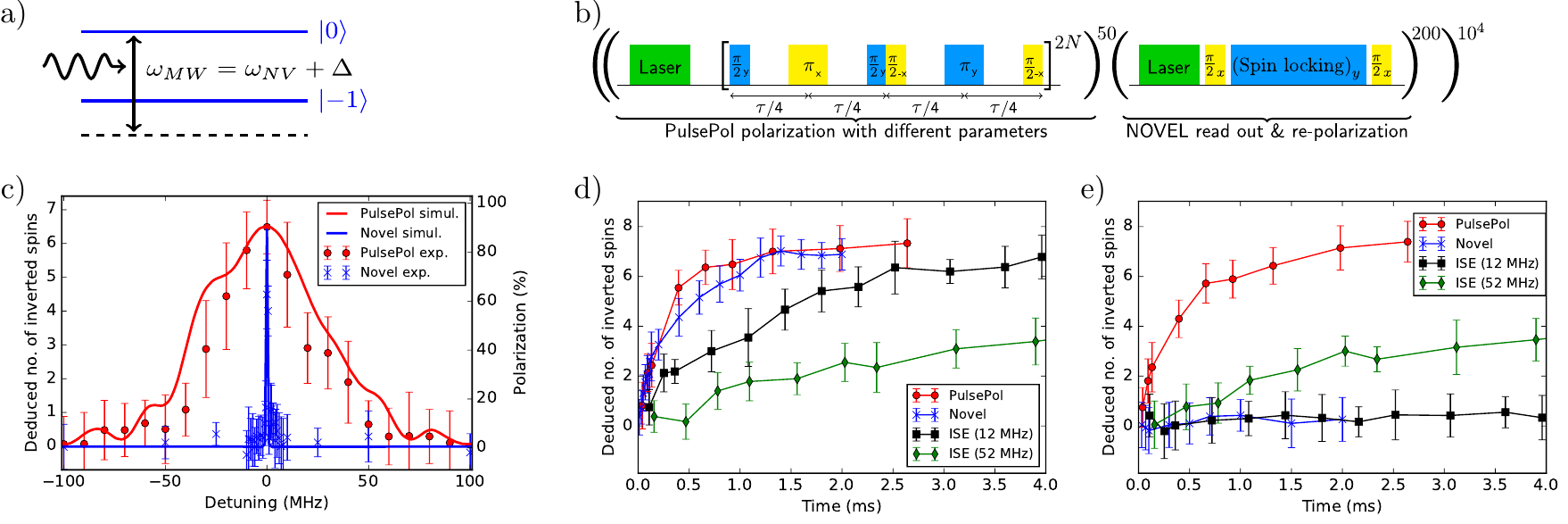}

\caption{(a) Probing robustness to detuning on a single NV by detuning the microwave frequency
from the NV $\ket{m_s=0} \leftrightarrow \ket{m_s=-1}$transition.
		(b) Polarisation readout by polarisation inversion (PROPI) sequence used for detecting polarisation efficiency of PulsePol.
		(c) PROPI readout for different detunings $\Delta$ for PulsePol (red) and NOVEL (blue). The lines are the smoothed simulation result of a comparable nuclear spin bath with no free parameters. For more details see text.
	Polarisation buildup by consecutive polarisation transfers for NOVEL, PulsePol and ISE (d) for $\Delta = 0$ MHz, where all sequences, (e) $\Delta = \bene{(2\pi)}20$ MHz, where only PulsePol and very slow ISE are able to transfer polarisation.}
	\label{experiment}
	\label{experiment2}	
\end{figure*}

We find that the PulsePol polarisation protocol is robust for a wide range of detuning
and Rabi frequency errors, sufficient for overcome inhomogeneities and enhancing the
polarisation transfer efficiency in many applications (e.g. most radicals or electron
defects in a solid/glass). However, for overcoming the wide spectral range inherent in
some systems, including NV centres in nanodiamonds and photoexcited triplet states, a
wider range of detuning robustness is beneficial. PulsePol can be optimised further by
making use of broadband universal rotations that have been developed in NMR spectroscopy.

Composite pulses or numerically optimised pulses, especially those derived with optimal
control algorithms, have been used in NMR~\cite{kobzar2012exploring,skinner2012new} and more
recently in EPR ~\cite{spindler2012shaped} can be used to engineer the pulses for
a required amount of robustness to detuning and pulse strength. Symmetric phase pulses
\cite{shaka1987symmetric} (see supplementary material \cite{suppinfo}) or the BURBOP
pulses defined in Ref.\cite{kobzar2012exploring} can extend the spectral range robustness
of PulsePol by a factor of two, to around $\bene{(2\pi)}120 \;\text{MHz}$  for the same pulse
duration as those used in Fig.~\ref{PulsePol_sim}. Importantly, experimental effects such as
the resonator bandwidth can be included in the numerical optimisation and accommodated for
in the engineered pulses~\cite{spindler2012shaped}.

It is worth noting that unlike traditional methods in EPR/NMR employing short pulses for
transferring polarisation between spins \bene{(e.g. pulsed ENDOR~\cite{davies1974new,MimsENDOR}, INEPT~\cite{morris1979INEPT1,morris1980INEPT2})}, no rf pulse or nuclear
spin manipulation is required in PulsePol, making it much easier to implement experimentally
and mitigating issues due to nuclear rotations taking 3 orders of magnitude longer time.


\textit{Experimental implementation via optically polarised NV centres in diamond ---} The NV
centre \mbpold{with its electron spin has recently gained considerable} interest as a resource spin
for optical DNP, for polarising nuclear spins inside the diamond \cite{london2013,scheuer2016optically,
chen2015optical, alvarez2015local,king2015room,pagliero2014recursive,drake2015influence} and in
external molecules \cite{Chen2015resinc,acebal2018toward,hollenberg2017,abrams2014dynamic}.
The NV centre realises in its ground state an electronic spin triplet ($S=1$) which exhibits
a zero-field splitting of $D = \bene{(2\pi)} 2.87$ GHz which separates the $m_s=0$ state energetically
from the $m_s=\pm 1$ states\bene{, which are split by an applied magnetic field. One of them forms with $m_s=0$ the desired effective two level system shown in figure \ref{experiment}(a)}. Importantly, the NV spin can be optically polarised within 200 ns by a laser pulse that induces spin-selective relaxation into the $m_s = 0$
sub-level of the ground
state resulting in spin polarisation exceeding 92\%~\cite{waldherr2011dark}.

%

%


\mbpold{Using this setup, the NV centre can be used to polarise the surrounding nuclear spins and to
use the same centre to readout the polarisation using Polarisation Read Out by Polarisation Inversion
(PROPI) \cite{scheuer2017propi} \bene{with the sequence in figure \ref{experiment}(b)}. This allows
to probe the effectiveness of polarisation schemes on the level of a single NV centre,} providing an
experimental test-bed for polarisation schemes. Figure~\ref{experiment}(c) demonstrates the polarisation
efficiency of PulsePol~\cite{fn2} and its robustness to detuning compared to NOVEL. The robust polarisation
of PulsePol for about $\bene{(2\pi)}60$MHz spectral width can be clearly seen, and is in very good
agreement with the theoretical simulations with 5 nuclear spins~\cite{fn3}. One should mention that this spectral width is only limited by the power
of the MW (the larger the power, the larger the width).


Previously developed methods to compensate for the lack of robustness of NOVEL and the
solid effect were based on sweep-based schemes, such as ISE. However, performing such
sweep induces a trade-off between its robustness to detuning and the polarisation efficiency
(the polarisation efficiency is inversely proportional to sweep speed, thus larger sweep
range either decreases efficiency or increases sequence time), which is not present in
PulsePol. Figure~\ref{experiment2}(d) shows the comparison for the polarisation build-up
rate of PulsePol, NOVEL and ISE (experimental parameters: Rabi frequency $\Omega =\bene{(2\pi)}1.79$ MHz
and inverse sweep rate of 3 $\mathrm{\mu s/\bene{(2\pi)}MHz}$, near optimal for this NV's coupling to
the nuclear spin bath), under exact resonance conditions. For ISE \mbpold{across a} $\bene{(2\pi)}12$ MHz
bandwidth (\mbpold{the minimal range that is required to achieve polarisation transfer for
typical parameters} \cite{scheuer2017propi}) and \mbpold{across a $\bene{(2\pi)}52$ MHz bandwidth (which
achieves polarisation transfer across a $\bene{(2\pi)}40$ MHz spectral range) were chosen. For the
case of perfect resonance, the buildup rate for PulsePol and NOVEL are comparable~\cite{fn4},}
while for ISE($\bene{(2\pi)}12$MHz) and ISE($\bene{(2\pi)}52$MHz) the rate is significantly slower due to the quadratic
dependence on the interaction with the nuclear spins in the efficiency of the MW sweep.
Figure~\ref{experiment2}(e) shows a similar comparison with $\Delta = \bene{(2\pi)}20$ MHz detuning.
While the performance of PulsePol and ISE($\bene{(2\pi)}52$MHz) remain essentially unchanged, \mbpold{for
both, NOVEL and ISE($\bene{(2\pi)}12$MHz), the} NV spin no longer transfers polarisation to the surrounding
nuclear spin bath.

\begin{figure}
 	\center
 	\resizebox{.48 \textwidth}{!}{
\includegraphics[scale=1]{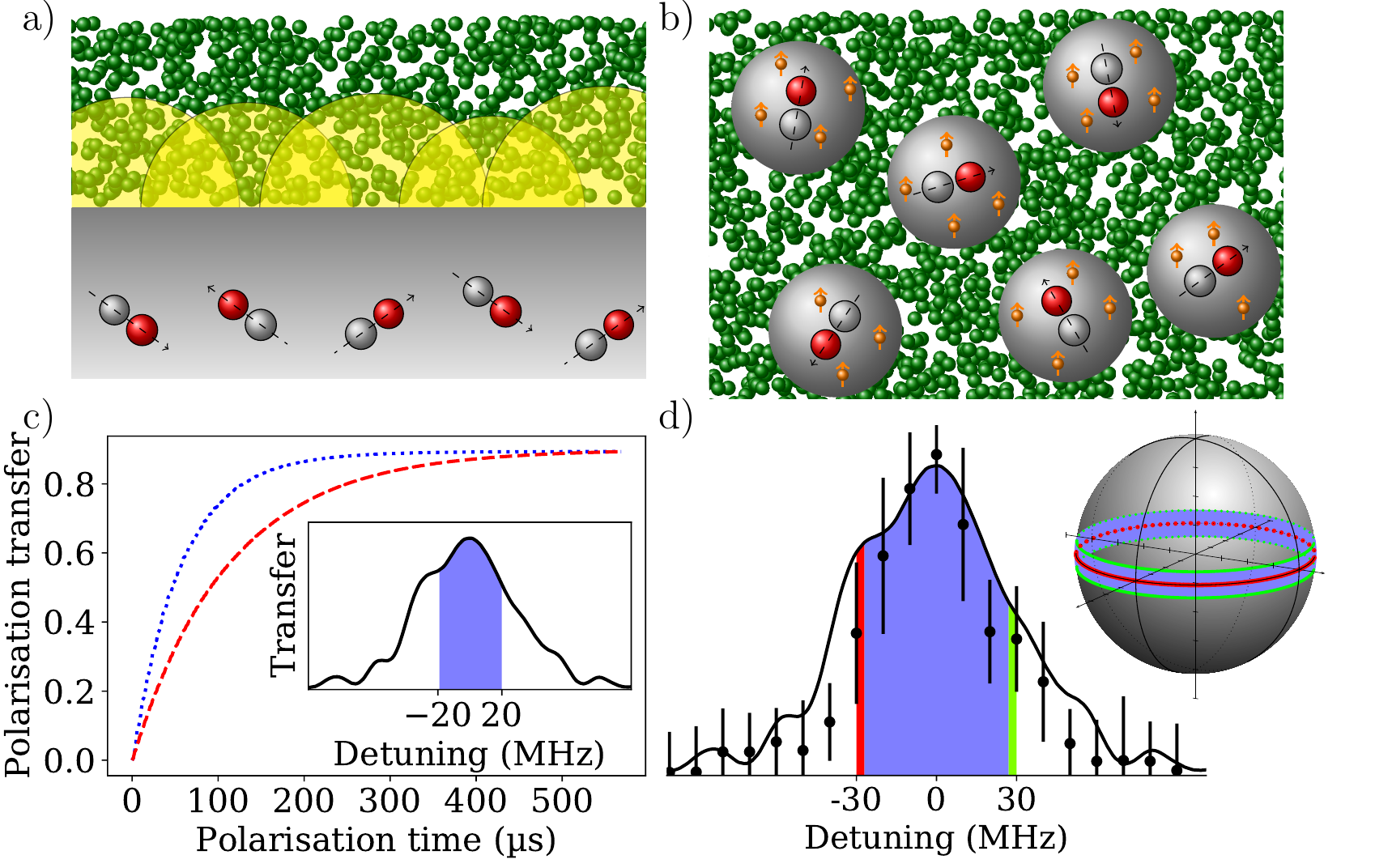}

}
 		\caption{(a) Illustration of shallow NV polarisation setup, where NV centres implanted $\sim3$ nm deep polarise diffusing molecules in a solution outside of the diamond and (b) nanodiamonds with randomly oriented NV centres, where the polarisation of $^{13}$C spins (orange) allows for the usage as MRI contrast agents.
 			(c) Polarisation transfer from near surface NV centres to a bath of 1200 nuclear spin bath, using the same Rabi and Larmor frequency and sequence as in figure 2(c) averaged over NV centres with transition frequencies drawn from a Gaussian distribution with $\bene{(2\pi)}20$MHz width. The simulation uses matrix product states \cite{schollwock2011density, mpnum} and a diffusion coefficient of $D = 1.4/2.8 \times 10^{-12} \mathrm{m^2/ s}$ (blue dotted/red dashed curve), resulting in  different correlation times and efficiencies.
 			The insert shows that the NV linewidth is well within the working range of the PulsePol protocol ($\Omega = \bene{(2\pi)}50$MHz).
 			(d) Based on the resilience to detunings $|\Delta| < \bene{(2\pi)}30$MHz for a Rabi frequency of $\Omega = \bene{(2\pi)}50$MHz, more than 11\% of the NV orientations in nanodiamonds (azimuthal angle between $90^\circ\pm 6.5^\circ$, as shown on the sphere) contribute to polarisation transfer.
 	 }
 	\label{Ish}
 \end{figure}

{
\textit{\mbpold{Discussion of} Applications ---} Important hyperpolarisation applications
with NV centres in diamonds where PulsePol is directly applicable are: (i) The polarisation
of nuclear spins in molecules external to the diamond with an ensemble of shallow NV centres
\cite{Chen2015resinc,acebal2018toward,hollenberg2017}, see Fig.~\ref{Ish}(a), where polarisation
transfer efficiency is the bottle-neck of the achieved polarisation. (ii) The hyperpolarisation
of nanodiamonds as MRI biomarkers. Nanodiamonds have been previously polarised at cryogenic temperatures (1-3K) and high magnetic fields~\cite{rej2015hyperpolarized,kwiatkowski2018direct, bretschneider2016potential, waddington2017phase}. However, optical polarisation via NV centres at room temperature offers several key advantages such as less complex and costly experimental setups and faster polarisation build-up time~\cite{chen2015optical}. For NV centre nanodiamond polarisation, an outstanding challenge is that the NV resonance can be significantly shifted due to
the random lattice orientation of the nanodiamond relatively to magnetic field orientation,
Fig.~\ref{Ish}(b) and (iii) the initialisation of quantum simulators based on 2 and 3D arrays
of nuclear spins \cite{cai2013}. Common to all these applications are the requirement of near
shallow NV centres which can exhibit between $\bene{(2\pi)}2$ and $\bene{(2\pi)}20$ MHz variance
in the NV resonance due to surface electric charges as well as interactions with P1 centres and
other surface defects \cite{RomachMU+15} (potentially more in dense ensembles). As the coupling between the shallow
NV spin and nuclear spin bath is very weak, even coherent polarisation transfer approaches the
NV relaxation time, and ISE will not dramatically increase the polarisation efficiency compared
with NOVEL. We simulate the polarisation transfer from the NV centre to diffusing molecules via
tensor-network methods for two diffusion coefficients in Fig.~\ref{Ish}(c), exploiting the
time-evolving block decimation (TEBD). Critically, PulsePol shows over 80\% polarisation
efficiency, whereas NOVEL (as well as ISE) is expected to perform on order of magnitude worse
for this parameter regime. The insert in the figure elucidates the reason for this high efficiency,
as the frequency detunings of most shallow NV centres are corrected by PulsePol. In the case
of nanodiamonds, due to the large zero-field splitting, the NV detuning can reach $\bene{(2\pi)}1.4$ GHz.
However, by choosing the microwave frequency near the NV resonance corresponding to 90$^\circ$
angle between the diamond axis and external field, over 11\% of the NV orientations can be
addressed within just $\bene{(2\pi)}60$ MHz~\cite{chen2015optical} which, in turn can be achieved by applying
PulsePol with $\bene{(2\pi)}49$MHz Rabi frequency (for comparison note that even in bulk diamond only 25\% NV
centres participate in polarisation dynamics due to 4 possible orientations in the lattice.}.
Due to Brownian rotation of the nanodiamonds in a solution, such wide NV addressability leads
to all nanodiamonds becoming hyperpolarised under reasonable conditions \cite{chen2015optical}.

\textit{Conclusion ---} In conclusion, we have introduced a \mbpold{framework that allows for the design
of highly robust and efficient pulsed DNP schemes that transfer polarisation from electron to nuclear
spins via pulse sequences that create an effective evolution that is described by a flip-flop Hamiltonian.
We presented a specific example sequence, PulsePol, which not only achieves polarisation transfer rates that
are similar to those for existing schemes when operated under ideal conditions, but also
affords a flexibility that confers a remarkable robustness to detuning, spectral width and pulse errors
because PulsePol does not need to satisfy a Hartmann-Hahn resonance for the microwave amplitude. We
underline the practical potential of PulsePol, with the experimentally demonstration of its efficiency
in transferring polarisation from an optically polarised NV centre in diamond to the surrounding $^{13}$C
nuclear spin bath over a range of over $\bene{(2\pi)}60$ MHz detuning, where this range is only limited by the MW
drive.} This efficient and robust polarisation transfer by PulsePol significantly enhances the potential
of hyperpolarisation of external molecules for nanoscale NMR or quantum simulators using near-surface NV centres and of nanodiamonds as
hyperpolarised MRI markers and as polarisation agents.

\mbpold{The PulsePol sequence and the framework within which it was derived possesses a considerable flexibility
which allows it to be used in a wide variety of DNP experiments,} potentially including those conducted via radicals
or at low temperatures. In addition to the advantages of robustness to microwave and magnetic field
inhomogeneity, the reduced sensitivity to the electron resonance frequency will enhance the polarisation
transfer in cases where hyperfine splitting or the unisotropic g-tensor broaden the electron spectral width.


\textit{Acknowledgements ---} We thank Jan F. Haase for useful discussions. This work was supported
by the EU project HYPERDIAMOND, the ERC Synergy grant BioQ, a PhD fellowship of the Integrated Center
for Quantum Science and Technology (IQST), the Humboldt Research Fellowship for Postdoctoral Researchers,
the DFG Collaborative Research Centre 1279 and  the  state  of  Baden-W{\"u}rttemberg  through bwHPC.

\textit{Methods ---} Experiments in diamond: The PROPI sequence used consists of two parts (see figure ~\ref{experiment}(b)) which are repeated $5\times10^4$ times. The experiment starts with a thermally polarised nuclear spin bath. With the first part of PROPI, nuclear spins are polarised into the $\ket{\uparrow }$ state with $50$ sequence cycles of PulsePol. Then with the second part of PROPI the nuclear spin bath is fully polarised  into the $\ket{\downarrow \downarrow \downarrow ...}$ direction with $200$ cycles of a well-known polarisation sequence; in this case, NOVEL with 10$\mu$s spin locking time and matched Hartmann-Hahn conditions is used. With this number of cycles a saturation of the polarisation transfer can be achieved. Hence, from the second repetition on, the first part of PROPI starts with a fully polarised nuclear spin bath ($\ket{\downarrow \downarrow \downarrow ...}$), which reverses nuclear spins into the opposite($\ket{\uparrow }$) direction up to a certain degree. The amount of transferred polarisation depends on the efficiency of the sequence to be tested, e.g. PulsePol, ISE or NOVEL.
The nuclear polarisation signal is observed during the re-polarisation part (second part) by monitoring the NV's fluorescence. A bright signal, originated from flip-flop processes between the NV electron spin and nuclear spins, saturates to a darker signal when the spin bath is reaching a completely polarised state. Hereby, the area below the saturation curve (first 100 out of $200$ cycles) gives a measure of nuclear polarisation.

To determine the behavior of PulsePol with regard to detuning errors, we have added a detuning
$\Delta = \omega_{1\leftrightarrow 0} - \omega_{MW}$ between the $\ket{m_s=0} \leftrightarrow
\ket{m_s=-1}$ transition frequency and the external microwave field frequency (see figure~\ref{experiment}(a)), using a drive of approx. $\Omega_0 =\bene{(2\pi)}49$ MHz Rabi frequency for the pulses. Additional experimental parameters: Rabi frequency $\Omega =\bene{(2\pi)}1.86$ MHz matching Hartmann-Hahn conditions and spin locking pulse of 10$\mu$s).

All experiments were performed at a magnetic field of approx. $1740\,$G and an alignment of the external field in respect to the NV axis to better than $1^{\circ}$. The microwave is applied to the NV through an electroplated stripline on diamond. The attenuation of this stripline is frequency dependent, such that the amplitude of the MW during PulsePol was corrected up to a deviation of the Rabi frequency of $<\pm5\%$. Potential depolarisation effects in the case of severely detuned pulses were checked and found to be negligible, see \cite{suppinfo}.

Polarisation of molecules via shallow NV centres: to simulate the surface NV ensemble interacting with diffusing spins we perform the following. The quantum system to be simulated comprises the electron spin of an NV-centre and multiple nuclear spins. The parallel and perpendicular coupling constants describing the NV-spin interaction are obtained numerically by simulating the diffusion of the $1200$ most strongly coupled spins, which accounts for two thirds of the total coupling strength. We simulate 100 different NV-spin systems with $\bene{(2\pi)}20$MHz standard deviation in the NV resonance frequency to account for the broad linewidth. During evolution under PulsePol, the entanglement between the different constituents is limited and can, therefore, be simulated efficiently via tensor-network methods. We exploit the time-evolving block decimation (TEBD) algorithm~\cite{schollwock2011density,mpnum} to perform the final simulations, as shown in Fig.~\ref{Ish}(c).

\bibliographystyle{plain}
\bibliographystyle{IEEEtran}

\begin{thebibliography}{10}
\bibitem{fn2}
	The resonance shift in the free evolution time due to hardware phase errors was
	checked and found to be near the optimal value.
\bibitem{fn3}
	A Gaussian filter was applied on the simulation to cancel very fast oscillating
	fluctuations due to the polarisation transfer resonance lines.
\bibitem{fn4}
	While the buildup for NOVEL should be a little faster, PulsePol encompasses the
	3 resonances due to the hyperfine interaction with the Nitrogen nuclear spin
	and therefore transfers more polarisation each cycle.
\end{thebibliography}

\newpage
\onecolumngrid
\newpage

\centerline{ \large \bf Supplementary Information}

\section{{The Hamiltonian of the system}}\label{Hamiltonian}

We consider a system of a single electron spin $\vec{S}$ ($S=\frac{1}{2}$) coupled to $N$ nuclear spins $\vec{I}^{(n)}$ (spin-1/2). The System is described by the Hamiltonian

\begin{equation} \label{H_SI}
H = \omega_S S_z + \sum\limits_{n=1}^{N} \omega_I I_z^{(n)} +  \sum\limits_{n=1}^{N} \vec{S} \mathcal{A} \vec{I}^{(n)} + 2 \Omega(t) S_x \cos(\omega_{MW} t + \varphi),
\end{equation}

where $\omega_S$($\omega_I$) denotes the electron (nuclear) Larmor frequency, $\mathcal{A}$ the hyperfine coupling tensor describing the interaction between electron and nuclear spins resulting from the magnetic dipole-dipole coupling, $\omega_{MW}$ the microwave frequency, $\varphi$ the microwave phase, and the Rabi frequency $\Omega(t)$ has the value $\Omega_0$ when the microwave is on, and $0$ otherwise.

In a rotating frame with respect to $\omega_S S_z$ and after applying the secular approximation, the Hamiltonian is
\begin{equation} \label{Hrot_SI}
H_{\text{int}} =  \Delta S_z +  \sum\limits_{n=1}^{N} \omega_I I_z^{(n)} +  \sum\limits_{n=1}^{N} S_z \vec{\mathcal{A}} \vec I^{(n)} + \Omega(t) (S_x \cos \varphi + S_y \sin \varphi),
\end{equation}
with $\Delta = \omega_S - \omega_{MW} $ denoting the detuning between the MW and electron Larmor frequency. In a nuclear spin basis such that $A_y = 0$ we obtain the basic Hamiltonian:
\begin{equation} \label{Hrot_fin}
H_{\text{int}} =  \Delta S_z +  \sum\limits_{n=1}^{N} \omega_I I_z^{(n)} +  \sum\limits_{n=1}^{N} S_z {\left(A_x I_x^{(n)} + A_z I_z^{(n)} \right)}+ \Omega(t) (S_x \cos \varphi + S_y \sin \varphi).
\end{equation}




According to this Hamiltonian, 
the free evolution operator for a time $\tau$ is
\begin{align}
U_{\text{free}}(\tau) &= \exp \left(-i \tau \left( \Delta S_z +  \sum\limits_{n=1}^{N} \omega_I I_z^{(n)} +  \sum\limits_{n=1}^{N} S_z \left(A_x I_x^{(n)} + A_z I_z^{(n)} \right) \right)\right) \nonumber
\\ &= \exp \left(-i \tau \left(\sum\limits_{n=1}^{N} \omega_I I_z^{(n)} +  \sum\limits_{n=1}^{N} S_z \left(A_x I_x^{(n)} + A_z I_z^{(n)} \right) \right)\right)  \exp \left(- i \tau \Delta S_z \right),\label{free2}
\end{align}
where the last equality holds as the $\Delta S_z$ term commutes with the other terms in the exponent.

The pulses have an additional Rabi term and are described by
\begin{equation}\label{pulses}
U_{\phi, \mathrm{\pm X/ \pm Y}} = \exp \left(-i \frac{\phi}{\Omega_0} \left( \Delta S_z +  \sum\limits_{n=1}^{N} \omega_I I_z^{(n)} +  \sum\limits_{n=1}^{N} S_z \left(A_x I_x^{(n)} + A_z I_z^{(n)} \pm \Omega_0 S_\mathrm{X/Y}    \right) \right)\right).
\end{equation}
As it is difficult to understand the dynamics from this description, the next section uses a model with only one nuclear spin to derive an effective Hamiltonian.

\section{The effective Hamiltonian of 
	PulsePol}\label{analytical}

The PulsePol sequence as shown in the text is given by 
\begin{equation}\label{SEQ_INTRO} 
\left[ 
\left( \frac{\pi}{2} \right)_{Y} 
\overset{\tau/4}{-\mathrel{\mkern-15mu}-\mathrel{\mkern-15mu}-}
\left( \pi \right)_{X} 
\overset{\tau/4}{-\mathrel{\mkern-15mu}-\mathrel{\mkern-15mu}-}
\left( \frac{\pi}{2} \right)_{Y} 
\left( \frac{\pi}{2} \right)_{-X} 
\overset{\tau/4}{-\mathrel{\mkern-15mu}-\mathrel{\mkern-15mu}-}
\left( \pi \right)_{Y} 
\overset{\tau/4}{-\mathrel{\mkern-15mu}-\mathrel{\mkern-15mu}-}
\left( \frac{\pi}{2} \right)_{-X} 
\right]^{2N}.
\end{equation}

where $(\phi)_{X, \pm Y}$ denote pulses around the X-/Y-axis  with duration $t' = \phi/\Omega$ and phase $\varphi = 0, \pm \pi/2$ and $\overset{\tau/4}{-\mathrel{\mkern-15mu}-\mathrel{\mkern-15mu}-}$ denotes a free evolution for a time $\tau/4$, which depends on the nuclear Larmor frequency. The basic sequence block is repeated 2$N$ times, where $N$ is a positive integer.

Here we express the spin operators in terms of the Pauli matrices $\sigma_i$ ($i \in \{x,y,z\}$): $S_i = \frac{1}{2} \sigma_i$. {According the PulsePol sequence in Eq. (\ref{SEQ_INTRO}), by using
	$
	U_{\frac{\pi}{2}, \text{X}} \begin{pmatrix}
	\sigma_x\\\sigma_y\\\sigma_z 
	\end{pmatrix} U_{\frac{\pi}{2}, \text{-X}}= \begin{pmatrix}
	\sigma_x\\\sigma_z\\-\sigma_y 
	\end{pmatrix}
	$
	and
	$
	U_{\frac{\pi}{2}, \text{Y}} \begin{pmatrix}
	\sigma_x\\\sigma_y\\\sigma_z 
	\end{pmatrix} U_{\frac{\pi}{2}, \text{-Y}}= \begin{pmatrix}
	-\sigma_z\\\sigma_y\\\sigma_x 
	\end{pmatrix}
	$, one can rewrite the evolution as}
\begin{equation}
U_{\text{cycle}}
= 
U_{\frac{\tau}{4}, \text{-Y}} 
U_{\frac{\tau}{4}, \text{Y}} 
U_{\frac{\tau}{4}, \text{-X}} 
U_{\frac{\tau}{4}, \text{X}} 
U_{\frac{\tau}{4}, \text{Y}} 
U_{\frac{\tau}{4}, \text{-Y}} 
U_{\frac{\tau}{4}, \text{X}} 
U_{\frac{\tau}{4}, \text{-X}} 
\end{equation}
where
\begin{equation}\label{UUs}
U_{\tau, \pm\text{X/}\pm\text{Y}} = \exp \left(-i \sum\limits_{n=1}^{N}\tau \left(\omega_I I_z^{(n)} \pm S_{x/y} (A_x I_x^{(n)} + A_z I_z^{(n)}) \right)  \right).
\end{equation}

This means the evolution can be expressed with an effective Hamiltonian
\begin{equation}\label{Ham03}
H_\text{eff}^{(1)} = \sum\limits_{n=1}^{N} \omega_I I_z^{(n)} + (-f_1(t)  S_x - f_2(t) S_y )(A_x I_x^{(n)} + A_z I_z^{(n)}) .
\end{equation}

The functions $f_{1/2}$ and how they emerge from the sequence are shown in figure \ref{fig:fourier}. They can be expressed as Fourier series:

\begin{figure}[t]
	\centering
	\includegraphics[width=.95\textwidth]{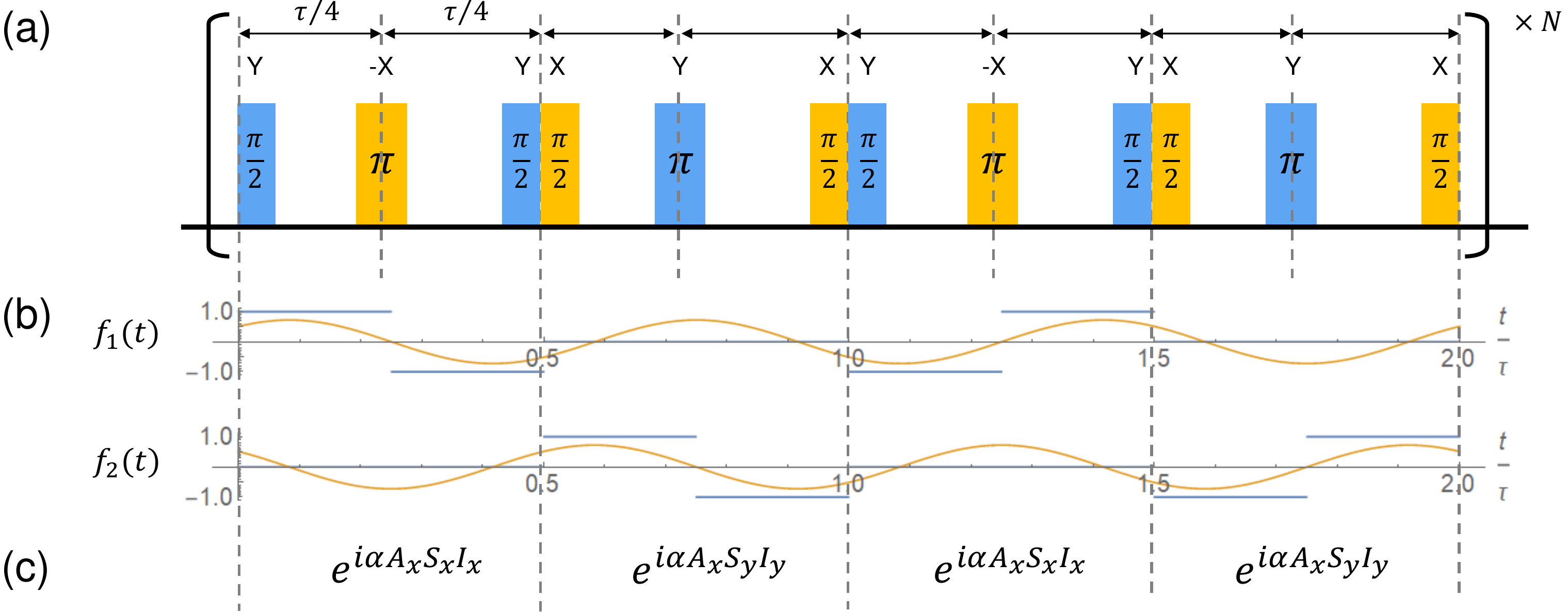}
	\caption{(a) Schematics of the MW pulse sequence on the electron spin for pulsed polarisation
		transfer to nuclear spins. Blue bars denote Y pulses and orange ones X or -X pulses. The sequence
		alternates between four sections. (b) The values of the modulation functions for the sections are
		$f_1(t)\sim \cos(3\pi t/\tau-\pi/4)$ and $f_2(t)\sim -\sin(3\pi t/\tau-\pi/4)$ for $S_x$ and $S_y$ (blue), and their dominant Fourier component (orange) (c) The pulse sequence produces
		$4$ distinct sections, each with an effective Hamiltonian in the rotating frame. The combination
		of a basis change for the electron spin in each section with the $\pi/2$ phase shift in the modulation
		function leads to the effective average Hamiltonian $S_{{x}} I_x +S_{{y}} I_y = 1/2(S_+I_- + S_-I_+)$
		which produces the polarisation transfer between electron spin and nuclei.}
	\label{fig:fourier}
\end{figure}

\begin{equation}
\begin{split}
f_1(t+k \cdot 2 \tau) &= \left\{\begin{array}{lr}
1, & \text{for } 0 \tau \leqslant t \leqslant \tau/4, 
\text{ } 5\tau/4 \leqslant t \leqslant 3 \tau/2\\
-1, & \text{for }  \tau/4 \leqslant t \leqslant \tau/2, 
\text{ } \tau \leqslant t \leqslant 5\tau/4\\
0, & \text{otherwise} 
\end{array}\right.  
\\&= \sum\limits_{n=0}^{\infty}  a_n^{(1)} \cos \frac{\pi n t}{\tau}
+b_n^{(1)} \sin \frac{\pi n t}{\tau}
\end{split}
\end{equation}

and
\begin{equation}\label{f2ser}
\begin{split}
f_2(t+k \cdot 2 \tau) &= f_1\left(t - \frac{\tau}{2}\right) = 
\sum\limits_{n=0}^{\infty}  a_n^{(2)} \cos \frac{\pi n t}{ \tau}
+b_n^{(2)} \sin \frac{\pi n t}{\tau}.
\end{split}
\end{equation}

The Fourier coefficients can be calculated as

\begin{align}
a_n^{(1)} &= \frac{2}{2 \tau}\int\limits_0^{8\tau} f_1(t) \cos \frac{\pi n t}{\tau} 
= \frac{1}{\pi n} \frac{1 - (-1)^n}{2} \left[ 4 \sin \frac{\pi n}{4} - 2 \sin \frac{\pi n}{2} \right].
\end{align}
This expression equals $0$ for every even value of $n$, the same holds for
\begin{align}
b_n^{(1)} &= \frac{2}{2 \tau}\int\limits_0^{8\tau} f_1(t) \sin \frac{\pi n t}{\tau} 
= \frac{1}{\pi n} \frac{1 - (-1)^n}{2} \left[- 4 \sin \frac{\pi n}{4} + 2  \right].
\end{align}

Due to the shift in the modulation functions \ref{f2ser} we get similar coefficients for $f_2$, 
$|a_n^{(2)}| = |b_n^{(1)}|$ and $|b_n^{(2)}| = |a_n^{(1)}|$.
{Notice that
	$
	a_1^{(1)} = - b_1^{(1)} = \frac{2}{\pi} \left(\sqrt{2}-1\right)
	$
	and
	$
	a_3^{(1)} =  b_3^{(1)} = \frac{2}{3 \pi} \left( \sqrt{2}+1  \right)
	$
	result in $|a_3^{(1)}|\approx 1.94$ $|a_1^{(1)}|$, which is the reason we choose the third order, namely $a_3^{(1)}$ and $ b_3^{(1)}$.
	{The related Fourier terms corresponding to $a_3^{(1)}$ and $ b_3^{(1)}$ are plotted in the upper figure \ref{fig:fourier}.
		This is achieved by choosing
		\begin{equation}
		\tau = 3 \frac{\pi}{\omega_I}.
		\end{equation}

		In a rotating frame with respect to $\omega_I I_z$, discarding all fast-rotating terms, we get
		
		\begin{align}\label{Ham03_YZ}
		H_{avg} &= -\sum\limits_{n=1}^{N}\frac{A_x}{2} a_3^{(1)} \left( S_x I_x^{(n)} +  S_y I_x^{(n)} -  S_x I_y^{(n)} +  S_y I_y^{(n)}  \right) \nonumber
		\\&= -\sum\limits_{n=1}^{N}\frac{A_x}{4} \alpha  \left(  \tilde S_- I_+^{(n)} +   \tilde S_+ I_-^{(n)}   \right) 
		\end{align}
	}
	with $\alpha = \sqrt{a_3^{(1)}+b_3^{(1)}} = \frac{2}{3 \pi} (2+\sqrt{2}) $. The basis change resulting from
	$
	\tilde S_x = ({S_x + S_y}){\sqrt 2}
	$
	and
	$
	\tilde S_y = ({-S_x + S_y}){\sqrt 2}
	$
	does not affect the polarisation dynamics as only the x-y-plane is rotated and the z-axis remains.

}

\section{Error Robustness}\label{Pol20der}
As a robust polarisation sequence PulsePol should meet the following criteria (i)-(iii) from the main text: \\
(i) It produces both $S_x$ and $S_y$ terms in the effective Hamiltonian with the modulation functions 
$f_1(t)$, $f_2(t)$, preferably by producing the phase change $f_2(t) = f_1(t+\pi/2)$. \\
(ii) The detuning 
errors accumulated during the free evolution need to be cancelled and 
preferably also decoupled from unwanted noise and fluctuations.\\
(iii) as pulses are not perfect, i.e. 
of a finite length and with detuning and Rabi frequency errors, the polarisation sequence should cancel 
such errors at least to the first order.\\
As criterion (i) is fulfilled as already shown in the previous section, we present the detailed proofs of the other criteria (ii) and (iii) as follows. 

Criterion (ii), the cancellation of detuning errors accumulated during the free evolution, follows from the anti-commutation relation of Pauli matrices
$[\sigma_{x/y}, \sigma_z]_+ = 0.$ This relation can be used to derive the property
$
\exp \left(-i {\tau} \Delta S_z \right) S_{x/y} 
= S_{x/y} \exp \left(+i {\tau} \Delta S_z \right).
$ As the $\Delta S_z$ term commutes with the rest of the free evolution, one can see the cancellation of the detuning errors accumulated during the free evolution around every perfect $\pi$ pulse in the evolution operator with the help of (\ref{free2})
\begin{align}\label{YerrorDet}
U_\text{free}(\tau) U_{\pi, \text{X/Y}} U_\text{free}(\tau) &= U_\text{free}(\tau) 2 S_{x/y} U_\text{free}(\tau)
\\& = e^{-i \tau (\omega_I + S_z A_x I_x)}  e^{-i \tau \Delta S_z} 2 S_{x/y} e^{-i \tau \Delta S_z} e^{-i \tau (\omega_I + S_z A_x I_x)}
\\&= U_\mathrm{free, \Delta = 0}(\tau) U_{\pi, \text{X/Y}} U_\mathrm{free, \Delta = 0}(\tau).
\end{align}

The cancellation happens in every part of the sequence, leading to complete cancellation of detuning errors for perfect pulses. This means that improving the pulses also corrects errors accumulated during the free evolution.

For criterion (iii), the cancellation of pulse errors to 1st order, is fulfilled.
Including a Rabi frequency error $\delta\Omega = \Omega_0 - \tilde \Omega_0$, the pulses with errors take the form

\begin{equation}\label{imp1}
U_{\theta, \pm X/Y} = \exp \left(-i  \frac{\theta}{\Omega} \left(\pm \tilde \Omega S_{x/y} + \Delta S_z  \right)  \right)
\end{equation}

{Introducing the definitions
	$
	\frac{\Delta}{\Omega} \equiv \epsilon k_1	
	$ and
	$
	\frac{\tilde \Omega}{\Omega} = 1 - \frac{\delta \Omega}{\Omega} \equiv 1 + \epsilon k_2,
	$ it is straightforward to show that in the evolution operator of one sequence block, neglecting the nuclear spins, the detuning and Rabi errors have an overall effect of
	
	\begin{align}\label{picancelrab}
	&	U_{\pi/2, -X} U_{\pi, Y} U_{\pi/2, -X} U_{\pi/2, Y} U_{\pi, X} U_{\pi/2, Y} U_{\pi/2, -X} U_{\pi, Y} U_{\pi/2, -X} U_{\pi/2, Y} U_{\pi, X} U_{\pi/2, Y} 
	\\& = - (\mathbb{1} + \epsilon^2 k_1^2 (-2i \sigma_z) + \epsilon^3 k_1^2 k_2 2 \pi i \sigma_z) + O(\epsilon^4).
	\end{align}
	
	This means in the pulses Rabi errors are cancelled up to second order, detuning errors in first order. Note that the second order in detuning errors is a z-rotation and can therefore be compensated with a tau-shift for a specific detuning value, as shown for phase errors later.
	This completes the proof that the PulsePol sequence fulfills all of the desired properties.} {Notice that  the PulsePol sequence is not the only option but the best sequence found.} It can be derived as the simplest sequence fulfilling the above criteria (i)-(iii).

{There are other possible sequences fulfilling the above properties, for example the PolXY sequence
	
	\begin{equation}\label{SEQUENCE1}
	\begin{split}
	\left( \frac{\pi}{2} \right)_{Y} 
	&\left[ 
	\overset{\tau/2}{-\mathrel{\mkern-15mu}-\mathrel{\mkern-15mu}-}
	\left( {\pi} \right)_{X} 
	\overset{\tau}{-\mathrel{\mkern-15mu}-\mathrel{\mkern-15mu}-}
	\left( {\pi} \right)_{Y} 
	\overset{\tau}{-\mathrel{\mkern-15mu}-\mathrel{\mkern-15mu}-}
	\left( {\pi} \right)_{X} 
	\overset{\tau}{-\mathrel{\mkern-15mu}-\mathrel{\mkern-15mu}-}
	\left( {\pi} \right)_{Y} 
	\overset{\tau/2}{-\mathrel{\mkern-15mu}-\mathrel{\mkern-15mu}-}
	\right.\\&\left.
	\left( \frac{\pi}{2} \right)_{X}
	\overset{\tau}{-\mathrel{\mkern-15mu}-\mathrel{\mkern-15mu}-}
	\left( {\pi} \right)_{Y} 
	\overset{\tau}{-\mathrel{\mkern-15mu}-\mathrel{\mkern-15mu}-}
	\left( {\pi} \right)_{X} 
	\overset{\tau}{-\mathrel{\mkern-15mu}-\mathrel{\mkern-15mu}-}
	\left( {\pi} \right)_{Y} 
	\overset{\tau}{-\mathrel{\mkern-15mu}-\mathrel{\mkern-15mu}-}
	\left( \frac{\pi}{2} \right)_{X}
	\right]^N 
	\left( \frac{\pi}{2} \right)_{-Y},
	\end{split}
	\end{equation}
	
	with a resonance for $\tau = n \pi/\omega_L$,
	but PulsePol has the best properties concerning error stability. 
	Furthermore PolXY and other sequences showed an undesired depolarisation behaviour for detuning values close to the Rabi frequency, which is significantly reduced for PolsePol as shown in section \ref{expres}.}

A sequence which has almost identical behaviour to PulsePol is

\begin{equation}\label{SEQ_INTRO2} 
\left[ 
\left( \frac{\pi}{2} \right)_{Y} 
\overset{\tau/4}{-\mathrel{\mkern-15mu}-\mathrel{\mkern-15mu}-}
\left( \pi \right)_{Y} 
\overset{\tau/4}{-\mathrel{\mkern-15mu}-\mathrel{\mkern-15mu}-}
\left( \frac{\pi}{2} \right)_{-Y} 
\left( \frac{\pi}{2} \right)_{X} 
\overset{\tau/4}{-\mathrel{\mkern-15mu}-\mathrel{\mkern-15mu}-}
\left( \pi \right)_{X} 
\overset{\tau/4}{-\mathrel{\mkern-15mu}-\mathrel{\mkern-15mu}-}
\left( \frac{\pi}{2} \right)_{-X} 
\right]^{2N}.
\end{equation}

Combining this with the actual PulsePol sequence leads to

\begin{align}\label{SEQ_INTRO3} 
\nonumber
&\left[\left[ 
\left( \frac{\pi}{2} \right)_{Y} 
\overset{\tau/4}{-\mathrel{\mkern-15mu}-\mathrel{\mkern-15mu}-}
\left( \pi \right)_{X} 
\overset{\tau/4}{-\mathrel{\mkern-15mu}-\mathrel{\mkern-15mu}-}
\left( \frac{\pi}{2} \right)_{Y} 
\left( \frac{\pi}{2} \right)_{-X} 
\overset{\tau/4}{-\mathrel{\mkern-15mu}-\mathrel{\mkern-15mu}-}
\left( \pi \right)_{Y} 
\overset{\tau/4}{-\mathrel{\mkern-15mu}-\mathrel{\mkern-15mu}-}
\left( \frac{\pi}{2} \right)_{-X} 
\right]^{2}\right.
\\&
\left. \left[ 
\left( \frac{\pi}{2} \right)_{Y} 
\overset{\tau/4}{-\mathrel{\mkern-15mu}-\mathrel{\mkern-15mu}-}
\left( \pi \right)_{Y} 
\overset{\tau/4}{-\mathrel{\mkern-15mu}-\mathrel{\mkern-15mu}-}
\left( \frac{\pi}{2} \right)_{-Y} 
\left( \frac{\pi}{2} \right)_{X} 
\overset{\tau/4}{-\mathrel{\mkern-15mu}-\mathrel{\mkern-15mu}-}
\left( \pi \right)_{X} 
\overset{\tau/4}{-\mathrel{\mkern-15mu}-\mathrel{\mkern-15mu}-}
\left( \frac{\pi}{2} \right)_{-X} 
\right]^{2}\right]^N
\end{align}
which behaves as PulsePol, but here no oscillations for changing detuning values are present.

The latter two sequences and PulsePol itself can be derived with the following steps:
\begin{enumerate}
	\item Criterion (i) suggests that the modulation functions of such a sequence of length $2\tau$ consists of four different parts, $f_1(t)=0$ in the second and fourth part represents a specific basis choice and 
	$f_2(t) = f_1\left(t-\tau/2\right)$ represents the correct phase difference.
	
	\item As detuning errors need to be cancelled according to criterion (ii), a relation similar to (\ref{YerrorDet}) 
	needs to be fulfilled, i.e. the sequence must consist of blocks like $\overset{t}{-\mathrel{\mkern-15mu}-\mathrel{\mkern-15mu}-}
	\left( \pi \right)_{\Phi} 
	\overset{t}{-\mathrel{\mkern-15mu}-\mathrel{\mkern-15mu}-}$. This, combined with 1., immediately leads to the modulation functions in figure \ref{fig:fourier}(b).
	\item The basis changes, criterion (i), require $\pi/2$-pulses before and after these blocks,
	the error cancellation (iii) determines the form to be
	$\left( \frac{\pi}{2} \right)_{\pm Y} 
	\overset{t}{-\mathrel{\mkern-15mu}-\mathrel{\mkern-15mu}-}
	\left( \pi \right)_{Y} 
	\overset{t}{-\mathrel{\mkern-15mu}-\mathrel{\mkern-15mu}-}
	\left( \frac{\pi}{2} \right)_{\mp Y} $ or
	$\left( \frac{\pi}{2} \right)_{\pm X} 
	\overset{t}{-\mathrel{\mkern-15mu}-\mathrel{\mkern-15mu}-}
	\left( \pi \right)_{Y} 
	\overset{t}{-\mathrel{\mkern-15mu}-\mathrel{\mkern-15mu}-}
	\left( \frac{\pi}{2} \right)_{\pm X} $
	up to rotations and basis changes.
\end{enumerate}

This leads to the PulsePol sequence (\ref{SEQ_INTRO}) and other options like (\ref{SEQ_INTRO2}), (\ref{SEQ_INTRO3}).


\section{Finite pulses}\label{finite}

In section \ref{analytical} pulses were assumed to be perfect. With a finite Rabi frequency they also take a finite time, during which all parts of the system evolve, increasing the total time needed for every sequence block. In order to compensate that, the evolution times need to be reduced. As a block contains free evolutions lasting $2 \tau$, eight $\pi/2$-pulses and four $\pi$-pulses, the free evolution time is

\begin{equation}\label{tau1}
\tau = \frac{\pi}{\omega_I} \times n - \frac{4}{2} t_\pi - \frac{8}{2} t_{\pi/2},
\end{equation}

where $t_\pi$ and $t_{\pi/2}$ denote the time needed for $\pi$- and $\pi/2$-pulses. In case of simple pulses, the expression reduces to

\begin{equation}
\tau = \frac{\pi}{\omega_I} \times n - \frac{4}{2} \frac{\pi}{\Omega_0} - \frac{8}{2} \frac{\pi}{2 \Omega_0} = \frac{\pi}{\omega_I} \times n - 4 \frac{\pi}{\Omega_0}.
\end{equation}

In case of different pulses like in section \ref{composite}, the correction term changes with the pulse times. In all cases it is important that the pulses do not reduce the free evolution time by a significant amount, simulations show that the pulse time should be less than 20\% of the free evolution time. To reach this regime even for long pulses, the parameter $n$ can be increased as described in the previous section.

\section{{Composite pulses}}\label{composite}

The stability with respect to both detuning and Rabi frequency errors can be increased with composite pulses, which are designed to correct those errors within the pulses. As our main goal is to get reliable polarisation transfer for a wide range of detunings, we focus on the correction of detuning errors over Rabi frequency errors here.

In the numerical simulations we used different sequences described in \cite{shaka1987symmetric}.
Best results were achieved with the pulses
\begin{equation}
90 = \overline{16}\hspace{0.2 cm}300\hspace{0.2 cm}\overline{266}\hspace{0.2 cm}54\hspace{0.2 cm}\overline{266}\hspace{0.2 cm}300\hspace{0.2 cm}\overline{16}
\end{equation}
for $\pi/2$ pulses
and
\begin{equation}\label{picomppulse}
180 = 325\hspace{0.2 cm}\overline{263}\hspace{0.2 cm}56\hspace{0.2 cm}\overline{263}\hspace{0.2 cm}325
\end{equation}
for $\pi$ pulses.

The composite pulses mentioned above rotate around angles of $1218/180 \times \pi$  and $1232/180 \times \pi$, respectively, which means according to equation (\ref{tau1}) that for PulsePol
\begin{equation}
\tau = \frac{\pi}{\omega_I} \times l - \left( 2 \frac{1232}{180} + 4\frac{1218}{180}    \right)\frac{\pi}{\Omega_0}.
\end{equation}

\begin{figure}[h]
	\centering
	\includegraphics[width=.5\textwidth]{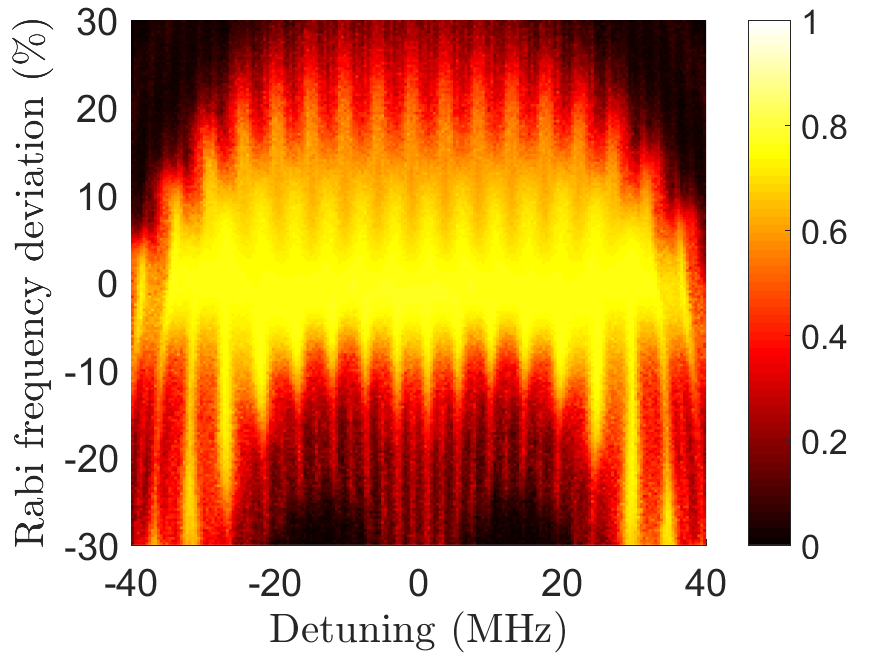}

	\caption{
		Error-resistance (polarisation transfer vs. $\Delta$ and $\delta\Omega/\Omega_0$) for PulsePol with composite pulses, parameters as in Fig. 2c in the main text: $\omega_I=2\;\text{MHz}$, $A_x=0.03\;\text{MHz}$ and $\Omega_0 = 50 \;\text{MHz}$. Here the $l=5$ resonance was used.
	}\label{fig:Cmp1}
\end{figure}

The resulting error robustness is shown in figure \ref{fig:Cmp1}.
The composite pulses allow for significantly larger detuning (For $\Omega = 50$ MHz more than $\pm40$ MHz) than in case of shorter pulses considered. The results aren't improved by choosing even longer composite pulses, because the pulse duration should not take a large fraction of the free evolution time.

%
%
%
%
%

\section{{The effect of phase errors}}

\begin{figure}[h]
	\centering
	\includegraphics[width=.65\textwidth]{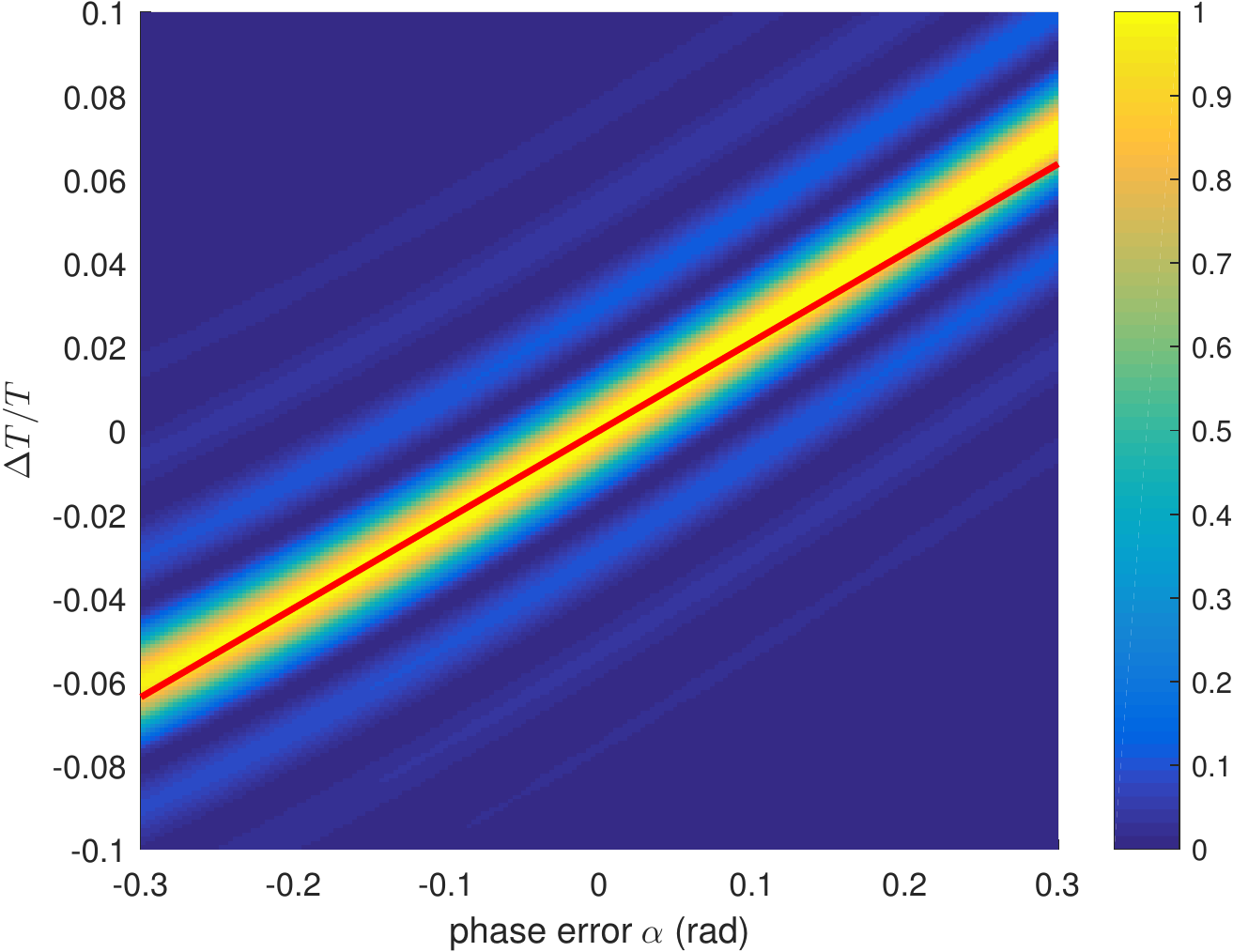}

	\caption{
		Simulated and calculated (red line, first order) dependence of the resonance shift on the error $\alpha$. A high function value (yellow) indicated that the protocol works well for the corresponding parameters.
	}\label{fig:TS1}
\end{figure}

In figure \ref{fig:TS1} one can see the analytically calculated (first order) and the simulated dependence of the shift in resonance depending on a phase error $\alpha$: In the PulsePol sequence, we assume every $\pi/2$ pulse that follows another pulse without a free evolution in between has a shifted phase

\begin{equation}\label{SEQ_INTRO_shift} 
\left[ 
\left( \frac{\pi}{2} \right)_{\overline Y} 
\overset{\tau/4}{-\mathrel{\mkern-15mu}-\mathrel{\mkern-15mu}-}
\left( \pi \right)_{X} 
\overset{\tau/4}{-\mathrel{\mkern-15mu}-\mathrel{\mkern-15mu}-}
\left( \frac{\pi}{2} \right)_{Y} 
\left( \frac{\pi}{2} \right)_{\overline{-X} }
\overset{\tau/4}{-\mathrel{\mkern-15mu}-\mathrel{\mkern-15mu}-}
\left( \pi \right)_{Y} 
\overset{\tau/4}{-\mathrel{\mkern-15mu}-\mathrel{\mkern-15mu}-}
\left( \frac{\pi}{2} \right)_{-X} 
\right]^{2N}.
\end{equation}

where
$\overline Y = Y \cos \alpha - X \sin \alpha
$ and
$
\overline{-X} = -X \cos \alpha + Y \sin \alpha.
$

This description has the same effect as standard phase errors, we choose it as in the experiment the shift in resonance condition originates from this effect.
For $\tau=0$ in first order of $\alpha$, the overall effect of this on the NV is described by

\begin{align}
&(	U_{\pi/2, -X} U_{\pi, Y} U_{\pi/2, \overline{-X}} U_{\pi/2, Y} U_{\pi, X} U_{\pi/2, \overline Y})^2\\
&=\left(	U_{\pi/2, -X} U_{\pi, Y} \left(U_{\pi/2, -X} + \frac{\alpha}{\sqrt{2}} U_{\pi, Y}\right)U_{\pi/2, Y} U_{\pi, X} \left(U_{\pi/2, Y} + \frac{\alpha}{\sqrt{2}} U_{\pi, -X}\right)\right)^2 + O(\alpha)^2\\
&=\exp\left(-4 i \alpha S_z\right) + O(\alpha)^2
\end{align}

This means the NV basis is rotated by $-2 \alpha$ in the x-y-plane during the time $2 \tau$. After a rotation of $-\pi$, the modulation functions are inverted. This corresponds to a shift of half a period in the modulation functions, which corresponds to a time 
\begin{equation}
\Delta T = \frac{1}{2 n} 2 \tau
\end{equation}
where $n$ is the (odd) resonance condition chosen. 

Assuming the sequence needs $M$ cycles for a $\pi$ shift, we can determine the relation between the resonance shift and the phase error

\begin{equation}
\frac{\Delta T/T}{\alpha} = \frac{\tau/n \times1/(2 \tau M)}{\pi/(4 M)} = \frac{2}{\pi n}.
\end{equation}

The corresponding line is plotted in figure \ref{fig:TS1} for $n=3$. Especially around $\alpha = 0$, where higher orders are negligible, it fits the simulation very well.

Note that this is a continuous process of shifting the modulation functions and therefore has a direction as the modulation functions are shifted by $\pi/2$. For $n=5$ the shift is in the opposite direction compared to $n=3$. For normal XY-sequences such a shift cannot induce a resonance shift, as only one modulation function does not allow for a direction.

\begin{figure}[h]
	\centering
	\includegraphics[width=.65\textwidth]{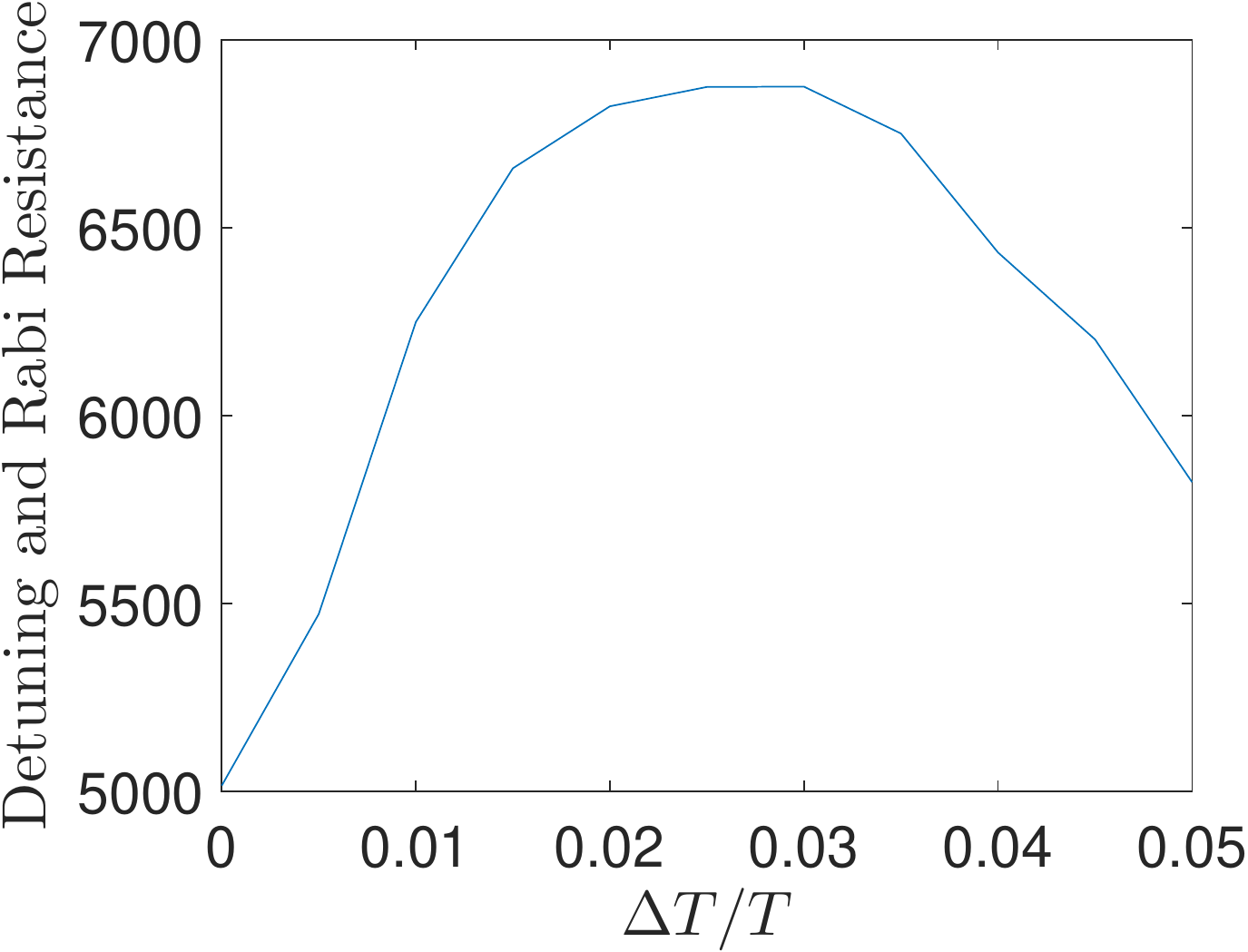}       
	
	\caption{
		Error resistance (arb. u.) of PulsePol depending on the resonance shift. }\label{fig:TS2}
\end{figure}

The major advantage of inducing such a resonance shift is that it allows to delete noise terms in the effective evolution.
Assuming an effective Hamiltonian
\begin{equation}
H_\text{eff} = \omega_I I_z + A_x S_\text{eff} I_x
\end{equation}

where
\begin{equation}
S_\text{eff} = f_1(t) \left(S_x + \epsilon_1 S_y + \epsilon_2 S_z \right) + f_2(t) \left(S_y + \epsilon_3 S_x + \epsilon_4 S_z \right),
\end{equation}

the errors $\epsilon_{1/3}$ would only slightly inhibit polarisation transfer, but the errors $\epsilon_{2/4}$ have a considerable impact. The shift in $\tau$ shifts the frequency of $f_{1/2}$ from the resonance and therefore decouples $S_z$ terms from the effective evolution. Only $S_{x/y}$ terms remain at the original resonance due to the rotation induced by $\alpha$.
Figure \ref{fig:TS2} shows that for $\Delta T/T \approx 2.5\%$ a considerably better resistance to Rabi and Detuning errors is achieved. The values result from an integration over heatmaps like in figure 2c in the main text (with a larger parameter space).
Note that everything described in this section is independent of the nuclear spin bath and therefore the resonance shift is equal for all NV centres in an ensemble. 

\section{Pulsed polarisation with NV-centers}\label{NVsec}

The basic Hamiltonian of the NV-center differs from the Hamiltonian discussed in the above sections by the fact that the electron spin $\vec S$ is now spin 1 and the zero field splitting $D$ gives rise to an additional term. We stay in the coordinate system where the $z$-axis is parallel to the magnetic field. Then the Hamiltonian can be written as

\begin{equation}
H = D S_{\tilde z}^2 + \omega_S S_z + \omega_I I_z + \vec{S} \mathcal{A} \vec{I} + 2 \Omega_3(t) S_x \cos(\omega_{MW} t + \varphi),
\end{equation}

As the orientation of the NV centre might not be parallel to the magnetic field defining the energy basis, the zero-field splitting term is in a different basis $\tilde z$, which transforms as

\begin{equation}
S_{\tilde z} = S_z \cos \theta + S_x \sin \theta \cos \phi + S_y \sin \theta \sin \phi.
\end{equation}
In this section we assume $\theta$ to be small.
Going to the rotating frame with respect to $\omega_{MW} S_z$ like in the two-level case, we have
\begin{equation}
H = D \left( \cos^2 \theta S_z^2 + \sin^2 \theta S_x^2 \right) + (\omega_S-\omega_{MW}) S_z + \omega_I I_z + S_z \vec {\mathcal{A}} \vec{I} + \Omega_3(t) (S_x \cos \varphi + S_y \sin \varphi).
\end{equation}

after a secular approximation assuming $\theta$ is small $\theta \lessapprox 10^\circ$.
Here $S_x^2 = S_y^2$ was used and rotating terms with a factor $\sin \theta$ vanished due to the secular approximation.
To be close to resonance, we want the effective detuning $D(\theta) + \omega_S-\omega_{MW}$ to be small. In this case the energy difference to the third level is large and we get close to the two-level case.

Now we can consider the matrix representation of the energy terms 

where we choose 
$
\Delta = \omega_S - \omega_{MW} + D(\theta)
$
with the splitting $D(\theta)$ up to third order in $\theta$
\begin{equation}
D(\theta) = D \left(\cos^2\theta +  \frac{1}{2} \sin^2\theta - \sin^2\theta\right) = D\left(1 -  \frac{3}{2} \sin^2 \theta\right) = \frac{D}{4}\left(1+3\cos(2\theta)\right).
\end{equation}

As the third level is far detuned by $2D(\cos^2\theta + \frac{1}{2} \sin^2\theta)$ it is not relevant for the dynamics for small values of $\Delta$.
This means the sequence now has the same effect as in the two-level case, with a two level Rabi frequency of $\Omega(t) = 1/\sqrt{2}\Omega_3(t)$. Similar calculations can be done for arbitrary values of $\theta$.

\section{Depolarisation behaviour}\label{expres}

\begin{figure}[h]
	\centering

	\includegraphics[width=.45\textwidth]{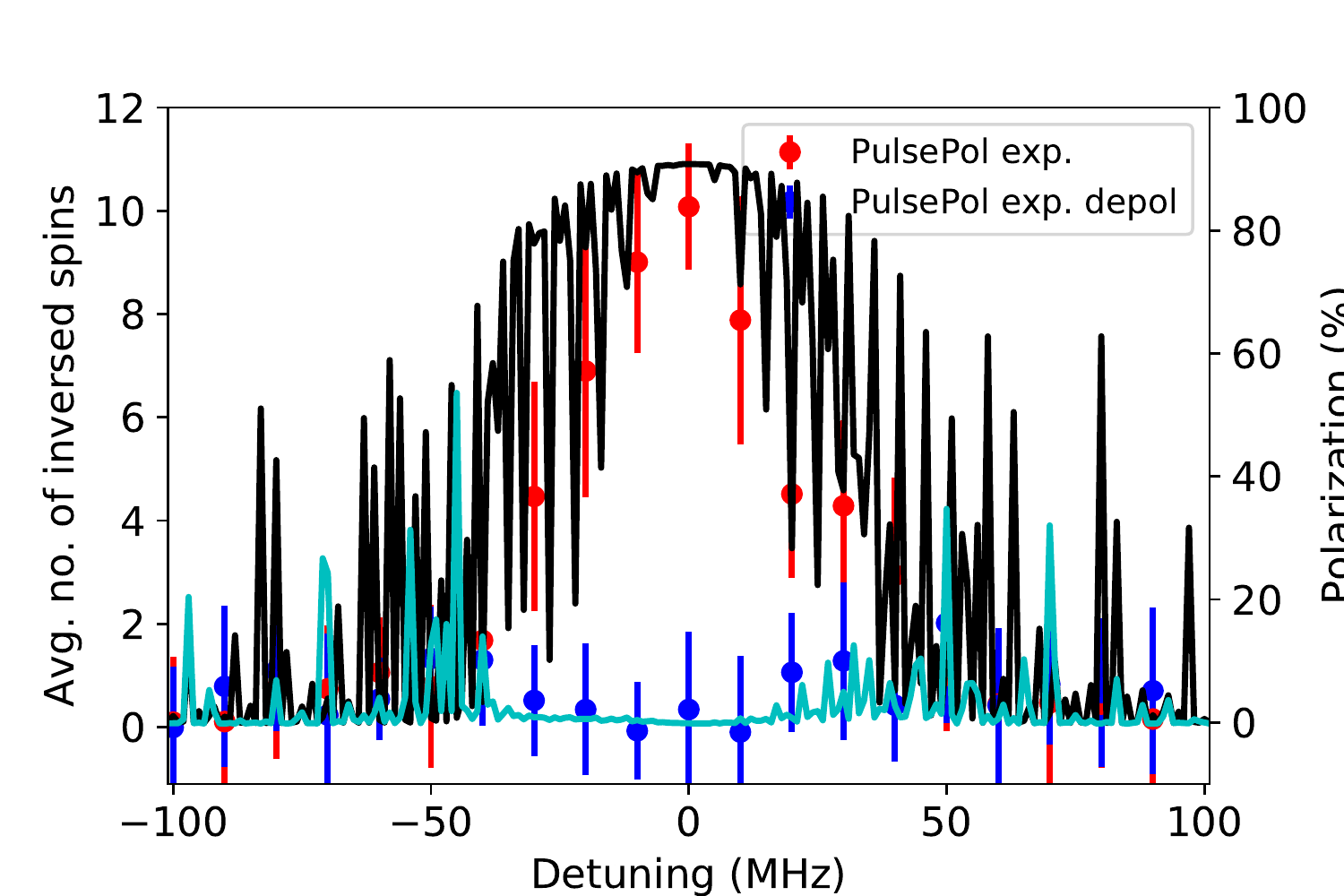}  
	\includegraphics[width=.45\textwidth]{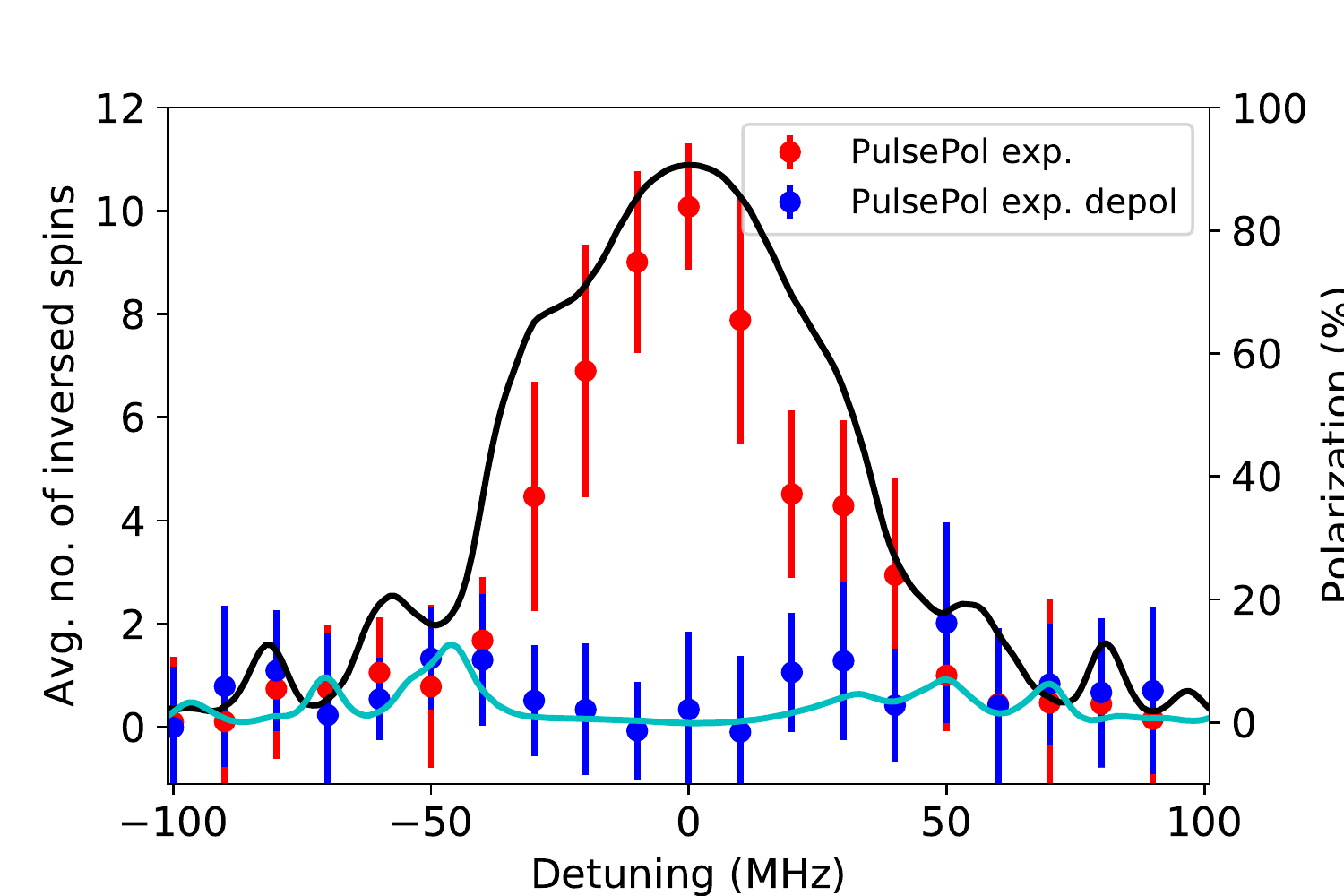}      
	
	\caption{
		Simulation results (left: raw data, right: smoothed) compared to the experimental data for polarisation buildup (red/black) and polarisation destruction (blue)}\label{fig:exp}
\end{figure}

The numerical results that are compared with experimental data in figure 3 in the main text were obtained by smoothing the actual data that were oscillating fast due to resonances of the free evolution detuning term. Figure \ref{fig:exp} also shows the unsmoothed data. Another important result presented in this figure is the depolarisation (blue): unlike other investigated sequences (like PolXY), PulsePol does not destroy a significant amount of polarisation for any detuning when applied in the other direction, i.e. to further polarise an already polarised bath.

This was tested with PROPI by reversing the polarisation direction during the read out (second part) from $\ket{\downarrow }$ to $\ket{\uparrow }$.

\end{document}